\documentclass[article]{article}
\usepackage{graphicx}
\usepackage{epstopdf}
\usepackage{amsmath}
\usepackage{amssymb}
\usepackage{mathrsfs}
\usepackage[numbers,sort&compress]{natbib}
\usepackage{amsthm}
\usepackage{pgfplots}
\usepackage{xcolor}
\usepackage{array}
\usepackage{float}
\usepackage{appendix}

\numberwithin{equation}{section}

\begin{document}

\baselineskip=18pt

\title{    Inverse scattering transformation for the    Sasa-Satsuma equation with non-vanishing    boundary conditions  }
\author{Li-Li Wen$^1$,  En-Gui Fan$^1$\thanks{\ Corresponding author and email address: faneg@fudan.edu.cn } }
\footnotetext[1]{ \  School of Mathematical Sciences, Fudan University, Shanghai 200433, P.R. China.}

\date{}

\maketitle
\begin{abstract}
\baselineskip=18pt
 We concentrate on inverse scattering transformation for  the Sasa-Satsuma equation with $3\times 3$ matrix spectral and nonzero boundary condition in this article.
 To circumvent multi valuedness  of   eigenvalues, we introduce a suitable two-sheet Riemann surface to map the original spectral parameter $k$ into a single-valued parameter $z$.
	The    analyticity   of the Jost  eigenfunctions and scattering coefficients  of Lax pair for the SS equation  are   analyzed in details.
  According to  the analyticity of  eigenfunctions and scattering coefficients,
 the $z$-complex plane  is divided into four analytic regions $D_j, \ j=1, 2, 3, 4$.  Since the second column of Jost eigenfunctions is  analytic    in $D_{j}, \ j=1, 2, 3, 4$,
 but in upper-half or lower-half plane,   we  introduce   certain  auxiliary eigenfunctions which are necessary for deriving the analytic eigenfunctions in $D_{j}$.
 We find that  for the  eigenfunctions, scattering coefficients and the auxiliary eigenfunctions  all possess three kinds of  symmetries,
 which characterize  the distribution  of discrete spectrum.
  The asymptotic behaviors of   eigenfunctions, auxiliary eigenfunctions and scattering coefficients  are also systematically derived.
 Then  a  matrix Riemann-Hilbert  problem with four kind jump conditions
  associated with the problem of nonzero asymptotic boundary conditions is established, from which
 $N$-soliton solutions is obtained via   the corresponding reconstruction formulae.
   The reflectionless soliton solutions are explicitly given.  As application of the $N$-soliton formula,   we  present  three kinds  of  single-soliton solutions
  according to   the distribution  of discrete spectrum.
\end{abstract}

\section{Introduction}

The nonlinear Schr$\ddot{o}$dinger equation
\begin{align}
&iq_{t}+\frac{1}{2}q_{xx}+|q|^{2}q=0\label{nls}
\end{align}
is one of the most important integrable equations appearing in various physical systems such as plasma physics, solid-state physics nonlinear optics and so on.
The  initial value problem   of  the NLS equation (\ref{nls})    was   solved by the inverse scattering transformation  (IST) method \cite{va1971}.  Hasegawa and Tappert found   the possibility of soliton propagation in optical fibers and showed the stability by numerical computations \cite{af1973}. In 1980,  Mollenauer, Stolen and Gordon observing the soliton propagation experimentally \cite{lr1980}. However, by the advancement of experiment accuracy,  several phenomena which can not be explained by the classical equation (\ref{nls}) have been observed in experimentally works. In order to explain these phenomena, Kodama and Hasegawa proposed a higher-order nonlinear Schr$\ddot{o}$dinger equation \cite{yk1985,ya9187}
\begin{equation}
iq_{t}+\frac{1}{2}q_{xx}+|q|^{2}q+i\varepsilon\{\beta_{1}q_{xxx}+\beta_{2}|q|^{2}q_{x}+\beta_{3}q(|q|^{2})_{x}\}=0, \quad \beta_{j}\in\mathbb{R},\label{hnls}
\end{equation}
which is   completely integrable  for  special  parameters    $\beta_1, \beta_2$ and $\beta_3$ \cite{da1983,hh1979,rh1973,mw1983}.
For the choice
$\beta_{1}:\beta_{2}:\beta_{3}=1:6:3$,  the equation (\ref{hnls}) reduced  to  \cite{ns1991}
\begin{equation}
iq_{t}+\frac{1}{2}q_{xx}+|q|^{2}q+i\varepsilon\{q_{xxx}+6|q|^{2}q_{x}+3q(|q|^{2})_{x}\}=0.\label{rhnls}
\end{equation}
If making    a transformation   \cite{ns1991}
\begin{subequations}
\begin{align}
&q(x,t)  \rightarrow q(x,t)e^{ \frac{i}{6\varepsilon}(x-\frac{t}{18\varepsilon})},  \ \  t\rightarrow t,\quad  x  \rightarrow  x+ \frac{t}{12\varepsilon}.\nonumber
\end{align}
\end{subequations}
  the equation (\ref{rhnls})  reduce to  the well-known Sasa-Satsuma (SS) equation
\begin{equation}
q_{t}+\varepsilon\{q_{xxx}+6|q|^{2}q_{x}+3q(|q|^{2})_{x}\}=0,\label{SS}
\end{equation}
 which   can be used to  describe the nonlinear phenomena in many situations, such as pulse propagation in optical fibers and deep ocean waves \cite{sh2013,na2015,jm2014,sg1999,cg2003}.
It was shown that the  SS equation  (\ref{SS})  is completely integrable and possesses  $3\times 3$  matrix spectral problem \cite{ns1991}.
  In recent years,  there has been  much work on the  Sasa-Satsuma equation.  For example,   bilinearization and multisoliton solutions for the SS  equation    \cite{cg2003},
 N-soliton solutions for  the   SS  equation  with  zero boundary condition by  IST method \cite{yj2009},      initial-boundary problems of  the  SS   equation    on   the half-line by Fokas uniformed method \cite{xj20131},
 N-soliton solutions for  the  SS   equation   with zero boundary condition by  Riemann-Hilbert approach \cite{wj2017},    long-time asymptotic of the SS equation  by  Deift-Zhou steepest descent method
 \cite{lgx2018},   high-order soliton solutions for  the  SS   equation   with zero boundary condition
 by  Riemann-Hilbert approach \cite{yc2019},    binary Darboux transformation  of  the   SS  equation    \cite{zh2017},    Bright-soliton and optical soliton   etc    \cite{lx2014,kn1998}.

In this article  we  are interested in  the   IST  for the SS  equation ($\varepsilon=1$)  by means of  Riemann-Hilbert (RH)  method  with the following nonzero boundary condition
\begin{equation}
\lim_{x\rightarrow\pm\infty}\mathbf{q}(x,t)=\mathbf{q}_{\pm}=\mathbf{q}_{0}e^{i\theta_{\pm}}\label{condition}
\end{equation}
where $\mathbf{q}$ and $\mathbf{q}_{0}$ are the two-component vectors, $\parallel\cdot\parallel$ is the standard Euclidean norm, $\theta_{\pm}$ are real numbers and $\parallel\mathbf{q}_{0}\parallel^{2}=q_{0}^{2}$.
Because  the SS equation   admits   a   $3\times 3$ matrix spectral problem,  which causes
   the  analyticity, symmetries and  asymptotic  to be  more complicated  than those of   $2\times 2$ matrix spectral problem.
 The auxiliary eigenfunctions  are also   necessary  to construct a matrix Riemann-Hilbert problem.

The  IST  method,  first discovered  by Gardner,  Green,
 Kruskal   and Miura, is  one of the  most powerful tool to  investigate solitons of  nonlinear models   \cite{csg1967,cs1974}.
 The  IST  for the focusing Schrodinger equation  with zero boundary conditions   was first developed by Zakharov
 and Shabat \cite{va1971},  later    for the defocusing case with nonzero boundary conditions  \cite{Zakharov1973}.
 The next important steps of the development of  IST method   is   the RH method  which as the  modern version of IST was   established  by Zakharov and Shabat  \cite{Zakharov1979}.
 It has since become clear that the    RH  method  is applicable to  investigate   exact  solutions   and  asymptotic  analysis of   solutions
for   a wide class of    integrable systems  \cite{at2004,xj2013,rj2014,hy2014,rb2007,at2016,yjk2010}.

The structure of this work is  as follows.   In section 2,  we investigate  the  spectral problem of  SS  equation (\ref{SS}) with the nonzero boundary condition (\ref{condition}) by introducing  a Riemann surface.
    To    construct  the  desired   RH problem,    we    investigate    the   analyticity  of   eigenfunctions,   auxiliary eigenfunctions  and  scattering coefficients.
 In  section 3,   we show  that all  eigenfunctions, scattering matrix and reflection coefficient   possess   three types symmetries.
  In  section 4, we  consider   the asymptotic of eigenfunctions, auxiliary eigenfunction and scattering coefficient.
 In section 5,   discuss the discrete spectrum and the residue conditions to analyze  poles for meromorphic matrices  appearing in  the RH problem.
 In   section 6,   we  present  a generalized RH problem with four jump matrixes, from  which we find a  reconstruction  formula   between solution of the  SS  equation  and the RH  problem.
   In section 7,   in  reflectionless case,  we discuss  solvability of the RH problem,  from which the  $N$-soliton solutions of the SS equation are  obtained.
    In section 8, as an illustrate examples,   we explicitly  obtain   three  kinds of one-soliton solutions according to different distribution of the spectrum,
    and their dynamical features   are analyzed.    Some proof process are provided in  Appendix.

\section{Spectral Analysis}
\subsection{Lax Pair}
The Sasa-Satsuma equation (\ref{SS}) admits  $3\times3$ matrix spectral problem \cite{ns1991}
\begin{equation}
\Phi_{x}=X\Phi,\ \  \Phi_{t}=T\Phi,\label{laxpair}
\end{equation}
where
\begin{eqnarray}
\begin{split}
&X=-ik\Lambda+Q,\\
&T=-4ik^{3}\Lambda+4k^{2}Q+2ik\Lambda(Q_{x}-Q^{2})+2Q^{3}-Q_{xx}+[Q_{x},Q],
\nonumber
\overline{}\end{split}
\end{eqnarray}
\begin{equation}
Q=\left(
\begin{array}{cc}
   \mathbf{0}& \mathbf{q} \\
 -\mathbf{q}^{\dag} & 0\\
\end{array}\right), \ \ \ \Lambda=\left(
\begin{array}{cc}
  \mathbf{I} & \mathbf{0}\\
  \mathbf{0}^{\dag} & -1\\
\end{array}\right),\nonumber
\end{equation}
in which some notations in (\ref{laxpair}) are  as follows:
\begin{itemize}
\item[$\blacktriangle$] {$k\in\mathbb{C}$ is the spectral parameter.}
\item[$\blacktriangle$]{vector function $\mathbf{q}=(q, \bar{q})^{T}$, the superscript $T$ denote the matrix transpose and $\bar{q}$ denotes the complex conjugation of $q$.}
\item[$\blacktriangle$]{the superscript $"\dag"$ represents the Hermitian of matrix.}
\item[$\blacktriangle$]{$\mathbf{I}$ is the identity matrix and $\mathbf{0}$ is the zero matrix or vector.}
\item[$\blacktriangle$]{eigenfunction $\Phi=(\Phi_{ij})_{3\times 3}, \quad (i,j=1,2,3)$.}
\end{itemize}

Under  the nonzero boundary condition (\ref{condition}),  we obtain the following asymptotic matrix spectral problem
\begin{equation}
\Phi_{x}=X_{\pm}\Phi, \quad \Phi_{t}=T_{\pm}\Phi,\label{lax2}
\end{equation}
where
\begin{subequations}
\begin{align}
&X_{\pm}=\lim_{x\rightarrow\pm\infty}X=-ik\Lambda+Q_{\pm},\nonumber\\
&T_{\pm}=\lim_{x\rightarrow\pm\infty}T=
-4ik^{3}\Lambda+4k^{2}Q_{\pm}-2ik\Lambda Q_{\pm}^{2}+2Q_{\pm}^{3},\nonumber\\
&Q_{\pm}=\lim_{x\rightarrow\pm\infty}Q=\left(
\begin{array}{cc}
   \mathbf{0}& \mathbf{q}_{\pm} \\
 -\mathbf{q}_{\pm}^{\dag} & 0\\
\end{array}\right).
\nonumber
\end{align}
\end{subequations}
The eigenvalues of matrices $X_{\pm}$ and $T_{\pm}$   are respectively
\[\begin{array}{ccc}
-ik, & i\lambda, & -i\lambda;\\
-4ik^{3}, & 2i\lambda(2k^{2}-q_{0}^{2}), & -2i\lambda(2k^{2}-q_{0}^{2}),
\end{array}\]
where
\begin{align}
\lambda^2 = k^{2}+q_{0}^{2} \label{brach}
\end{align}
which is a double-valued function for complex variable $k$.

Next, we derive the eigenvector of $X_{\pm}$ and $T_{\pm}$. For the convenient of expression, we firstly introduce the definition of orthogonal vector in the following Lemma.
\newtheorem{lemma}{Lemma}
\begin{lemma}\label{lemma1}
For any two-component complex-valued vector $\mathbf{l}=(l_{1},l_{2})^{T}$, one can define its orthogonal vector as
$\mathbf{l}^{\perp}=(l_{2},-l_{1})$, and holds the property $\mathbf{l}^{\perp}\mathbf{l}=  \mathbf{l}^T \mathbf{l}^{\perp}=0$.

\end{lemma}
It is obvious that  $X_{\pm}$ and $T_{\pm}$  possess  the   relation
$[X_{\pm},T_{\pm}]=0$. So $X_{\pm}$ and $T_{\pm}$ holds the common eigenvectors.
Through some calculations, one can obtain an invertible matrix $\Gamma_{\pm}$ satisfying  the following equation
\begin{equation}
X_{\pm}\Gamma_{\pm}=i\Gamma_{\pm}J, \quad  T_{\pm}\Gamma_{\pm}=i\Gamma_{\pm}\Omega,\nonumber
\end{equation}
where
\begin{subequations}
\begin{align}
&J=\mathrm{diag}\left(-\lambda,-k,\lambda\right),\nonumber\\
&\Omega=\mathrm{diag}\left(-2(2k^{2}-q_{0}^{2})\lambda,-4k^{3},2(2k^{2}-q_{0}^{2})\lambda\right),\nonumber\\
&\Gamma_{\pm}=\left(
\begin{array}{ccc}
   -\frac{\mathbf{q}_{\pm}}{q_{0}}& \frac{\mathbf{q}_{\pm}^{\perp}}{q_{0}}&-\frac{i\mathbf{q}_{\pm}}{k+\lambda} \\
 \frac{iq_{0}}{k+\lambda}& 0&1\\
\end{array}\right).
\nonumber\end{align}\end{subequations}

\subsection{Riemann Surface}
In the scalar case, $\lambda=\sqrt{k^{2}+q_{0}^{2}}$ is a branched function. The branch points are the values of $k=\pm iq_{0}$. We take the branch cut on the segment $[-iq_{0},iq_{0}]$. In this case, if we set
\begin{equation}
k+iq_{0}=r_{1}e^{i\alpha_{1}},\quad k-iq_{0}=r_{2}e^{i\alpha_{2}}, \quad -\frac{\pi}{2}<\alpha_{j}<\frac{3\pi}{2}, j=1,2,
\end{equation}
 we then  obtain two single value functions
\begin{equation}
\lambda=\pm\sqrt{r_{1}r_{2}}e^{\frac{i}{2}(\alpha_{1}+\alpha_{2})}.\nonumber
\end{equation}

We introduce  an  uniformization variable as
\begin{equation}
z=k+\lambda.\nonumber
\end{equation}
Substituting it into  ({\ref{brach}) gives two single-valued functions
\begin{equation}
k=\frac{1}{2}\left(z-\frac{q_{0}^{2}}{z}\right),\ \ \ \lambda=\frac{1}{2}\left(z+\frac{q_{0}^{2}}{z}\right).\nonumber
\end{equation}
Further  we can  show   the  following relations between the Riemann surface and the $k$-plane.
\begin{itemize}
\item[$\blacktriangle$] {The region where $\mathrm{Im}\lambda>0$  come from  the upper-half plane of the sheet-$\mathrm{I}$ and
the lower-half plane of the sheet-$\mathrm{II}$.  The region where $\mathrm{Im}\lambda<0$  come from  the upper-half plane of the sheet-$\mathrm{II}$ and the lower-half plane of the sheet-$\mathrm{I}$.}
\item[$\blacktriangle$]{On the sheet-$\mathrm{I}$, $z\rightarrow\infty$ as $k\rightarrow\infty$, and on the sheet-$\mathrm{II}$, $z\rightarrow 0$ as $k\rightarrow\infty$.}
\item[$\blacktriangle$]{The real $\lambda$ (real $k$) axes is mapped into the real $z$ axes.}
\item[$\blacktriangle$]{The branch cut $i[-q_{0},q_{0}]$ is mapped into the circle $C_{0}$ of the radius $q_{0}$ in $z$-plane.}
\item[$\blacktriangle$]{The sheet-$\mathrm{I}$ and sheet-$\mathrm{II}$, except for the branch cut, are mapped into the exterior and the interior of $C_{0}$, respectively.}
\end{itemize}
 The jump contour in the complex $z$-plane  is denoted by  $\Sigma=\mathbb{R}\cup C_{0}$.
The yellow  and white regions  in Fig.1  denote the analytic region  $D_{j} (j=1,2,3,4)$,  respectively
\begin{subequations}
\begin{align}
&D_{1}=\{z\in \mathbb{C}: |z|^{2}-q_{0}^{2}>0\cap\mathrm{Im}z>0\},\quad D_{2}=\{z\in \mathbb{C}: |z|^{2}-q_{0}^{2}>0\cap\mathrm{Im}z<0\},\nonumber\\
&D_{3}=\{z\in \mathbb{C}: |z|^{2}-q_{0}^{2}<0\cap\mathrm{Im}z<0\},\quad D_{4}=\{z\in \mathbb{C}: |z|^{2}-q_{0}^{2}<0\cap\mathrm{Im}z>0\},\nonumber
\end{align}
\end{subequations}
and $\bigcup\limits_{j=1}^{4}\bar{D}_{j}=\mathbb{C}$.
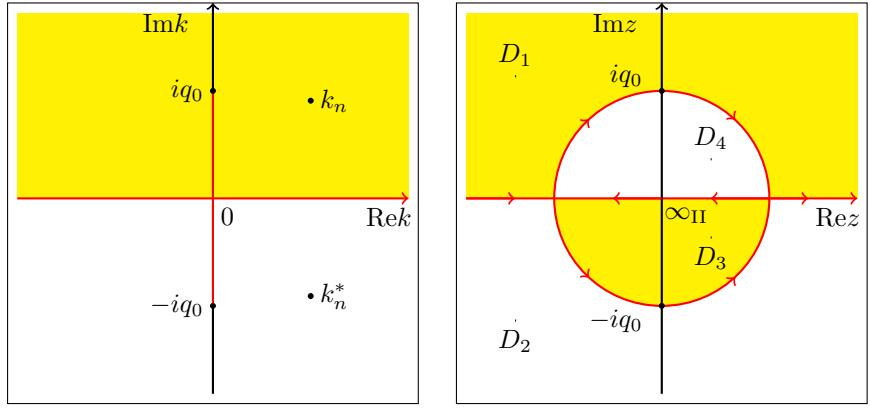
\begin{figure}
\begin{center}
\begin{tikzpicture}[scale=0.65]
\draw[ ](-4.2,-4.2)rectangle(4.2,4);
\path[fill=yellow] (-4,0)rectangle (4,3.8);
\draw[thick,red,->] (-4,0)--(4,0);
\draw[ ](0.3,0)node[below] {$0$} (3.6,0)node[below] {$\mathrm{Re}k$};
\draw[thick,red,-] (0,-2.2) -- (0,2.2);
\draw[thick,-](0,-4)--(0,-2.2);
\draw[thick,->](0,2.2)--(0,4);
\draw[ ](-0.3,3.6)node[left]{$\mathrm{Im}k$};
\draw [fill] (2,2) circle [radius=0.05] node[right] {$k_{n}$};
\draw [fill] (2,-2) circle [radius=0.05] node[right] {$k_{n}^{*}$};
\draw [fill] (0,2.2) circle [radius=0.05] node[left] {$iq_{0}$};
\draw [fill] (0,-2.2) circle [radius=0.05] node[left] {$-iq_{0}$};
\end{tikzpicture}
\quad
\begin{tikzpicture}[scale=0.65]
\draw[ ](-4.2,-4.2)rectangle(4.2,4);
\path[fill=yellow] (-4,0)rectangle (4,3.8);
\filldraw[white](0,0)--(2.2,0) arc (0:180:2.2);
\filldraw[yellow](0,0)--(-2.2,0) arc (180:360:2.2);
\draw[thick,red,->] (-4,0)--(4,0);
\draw[ ](0.5,0)node[below] {$\infty_{\mathrm{II}}$} (3.6,0)node[below] {$\mathrm{Re}z$};
\draw[thick,->](0,-4)--(0,4);
\draw[ ](-0.3,3.6)node[left]{$\mathrm{Im}z$};
\draw [red,thick] (0,0) circle [radius=2.2];
\draw [fill] (0,2.2) circle [radius=0.05];
\draw[ ](-0.2,2.5)node[left] {$iq_{0}$};
\draw [fill] (0,-2.2) circle [radius=0.05];
 \draw[ ](-0.2,-2.5)node[left] {$-iq_{0}$};
\draw[thick,red,->](-4,0)--(-3,0);
\draw[thick,red,->](0,0)--(-1,0);
\draw[thick,red,->](2,0)--(1,0);
\draw[thick,red,->](2,0)--(3,0);
\draw[thick,red,->](-1.5,1.61)--(-1.49,1.621);
\draw[thick,red,->](-1.5,-1.61)--(-1.49,-1.621);
\draw[thick,red,->](1.49,1.621)--(1.5,1.61);
\draw[thick,red,->](1.49,-1.621)--(1.5,-1.61);
\draw [thick,-] (-3,2.5) node[above] {$D_{1}$}--(-3,2.5);
\draw [thick,-] (-3,-2.5) node[below] {$D_{2}$}--(-3,-2.5);
\draw [thick,-] (1,-0.8) node[below] {$D_{3}$}--(1,-0.8);
\draw [thick,-] (1,0.8) node[above] {$D_{4}$}--(1,0.8);
\end{tikzpicture}
\end{center}
\caption{Left: The $k$-plane, showing the branch cut(red line), the branch points $\pm iq_{0}$, the region of $\mathrm{Im}k>0$(yellow) and $\mathrm{Im}k<0$(white); Right: The Riemann surface of $z$-plane, showing the analytic region $D_{+}$ $(\mathrm{yellow}, D_{1}, D_{3})$, $D_{-}$ $(\mathrm{white}, D_{2}, D_{4})$.}
\end{figure}

On the $z$ complex  plane,   we can rewrite the invertible matrix $\Gamma_{\pm}$ as
\begin{equation}
\Gamma_{\pm}=\left(
\begin{array}{ccc}
   -\frac{\mathbf{q}_{\pm}}{q_{0}}& \frac{\mathbf{q}_{\pm}^{\perp}}{q_{0}}&-\frac{i\mathbf{q}_{\pm}}{z} \\
 \frac{iq_{0}}{z}& 0&1\\
\end{array}\right),
\nonumber\end{equation}
moreover,
\begin{equation}
\begin{split}
&\det{\Gamma_{\pm}}=1+\frac{q_{0}^{2}}{z^{2}}\triangleq\gamma(z),\\
&\Gamma_{\pm}^{-1}=\frac{1}{\gamma}D(z)\left(
\begin{array}{cc}
   -\frac{\mathbf{q}_{\pm}^{\dagger}}{q_{0}}&-\frac{iq_{0}}{z} \\
 \frac{(\mathbf{q}_{\pm}^{\perp})^{\dagger}}{q_{0}}&0\\
 -\frac{i\mathbf{q}_{\pm}^{\dag}}{z}&1
\end{array}\right),\\
&D(z)=\mathrm{diag}\left\{1,\gamma(z),1\right\}.
\nonumber\end{split}\end{equation}

We also rewrite the Lax pair (\ref{laxpair}) as the polynomial form
\begin{equation}
\Phi_{\pm,x}=X_{\pm}\Phi_{\pm}+\Delta Q_{\pm}\Phi_{\pm},\ \ \
\Phi_{\pm,t}=T_{\pm}\Phi_{\pm}+\Delta\hat{Q}_{\pm}\Phi_{\pm},\label{lax}
\end{equation}
where
\begin{subequations}
\begin{align}
&\Delta Q_{\pm}=Q-Q_{\pm},\nonumber\\
&\Delta\hat{Q}_{\pm}=4k^{2}Q+2ik\Lambda(Q_{x}-Q^{2})+2Q^{3}-Q_{xx}+[Q_{x},Q]
-(4k^{2}Q_{\pm}-2ik\Lambda Q_{\pm}^{2}+2Q_{\pm}^{3}).\nonumber
\end{align}
\end{subequations}
\subsection{Analyticity}
In this subsection, we can define the Jost solutions $\Phi_{\pm}(x,t,z)$ as the simultaneous solutions of the Lax pair (\ref{laxpair}) with the boundary conditions
\begin{equation}
\Phi_{\pm}(x,t,z)\sim\Gamma_{\pm}(z)e^{i\Theta(x,t,z)},\quad\quad x\rightarrow \pm\infty, \label{phigamma}
\end{equation}
where
\begin{subequations}
\begin{align}
&\Theta(x,t,z)=Jx+\Omega t=\mathrm{diag}(\theta_{1},\theta_{2},-\theta_{1}),\nonumber\\
&\theta_{1}=-\lambda[x+2(2k^{2}-q_{0}^{2})t],\quad \theta_{2}=-k(x+4k^{2}t).\nonumber
\end{align}
\end{subequations}

To remove the asymptotic exponential oscillations, we introduce the following modified Jost eigenfunction
\begin{equation}
\mu_{\pm}(x,t,z)=\Phi_{\pm}(x,t,z)e^{-i\Theta(x,t,z)},\label{muphi}
\end{equation}
which implies that
\begin{equation}
\lim_{x\rightarrow\pm\infty}\mu_{\pm}(x,t,z)=\Gamma_{\pm}(x,t,z).
\end{equation}
After some simple calculation,  we derive the Lax pair of $\mu_{\pm}$
\begin{subequations}
\begin{align}
&(\Gamma_{\pm}^{-1}\mu_{\pm})_{x}=[iJ,\Gamma_{\pm}^{-1}\mu_{\pm}]+\Gamma_{\pm}^{-1}\Delta Q_{\pm}\mu_{\pm},\\
&(\Gamma_{\pm}^{-1}\mu_{\pm})_{t}=[i\Omega,\Gamma_{\pm}^{-1}\mu_{\pm}]+\Gamma_{\pm}^{-1}\Delta\hat{Q}_{\pm}\mu_{\pm},
\end{align}\label{mulax}
\end{subequations}
which  can be written in the full derivative form
\begin{equation}
\mathrm{d}\big[e^{-i\Theta(x,t,z)}\Gamma_{\pm}^{-1}\mu_{\pm}e^{i\Theta(x,t,z)}\big]=e^{-i\Theta(x,t,z)}[U_{1}\mathrm{d}x+U_{2}\mathrm{d}t]e^{i\Theta(x,t,z)},\label{fulld}
\end{equation}
where
\begin{equation}
U_{1}=\Gamma_{\pm}^{-1}\Delta Q_{\pm}(x,t)\mu_{\pm}(x,t,z),\quad U_{2}=\Gamma_{\pm}^{-1}\Delta\hat{Q}_{\pm}(x,t)\mu_{\pm}(x,t,z).\nonumber
\end{equation}
So we can come to the Jost integral equation from (\ref{fulld})
\begin{subequations}
\begin{align}
&\mu_{-}(x,t,z)=\Gamma_{-}+\int_{-\infty}^{x}\Gamma_{-}e^{i(x-y)J(z)}\Gamma_{-}^{-1}\Delta Q_{-}\mu_{-}e^{-i(x-y)J(z)}\mathrm{d}y,\\
&\mu_{+}(x,t,z)=\Gamma_{+}-\int_{x}^{+\infty}\Gamma_{+}e^{i(x-y)J(z)}\Gamma_{+}^{-1}\Delta Q_{+}\mu_{-}e^{-i(x-y)J(z)}\mathrm{d}y.
\end{align}
\end{subequations}
And we also derive the analyticity:
\newtheorem{thm}{\bf Theorem }[section]
\begin{thm}\label{thm1}
If $\mathbf{q}(\cdot,t)-\mathbf{q}_{+}\in\mathbf{L}^{1}(a,+\infty)$ or $\mathbf{q}(\cdot,t)-\mathbf{q}_{-}\in\mathbf{L}^{1}(-\infty,a)$ for any constant $a\in\mathbb{R}$, the following columns of $\mu_{+}(x,t,z)$ or $\mu_{-}(x,t,z)$ can be analytically extended onto the corresponding regions of the complex $z$-plane
\renewcommand\arraystretch{2}
\begin{table}[H]
\caption{the analytically of $\mu_{\pm}$}
\centering
\begin{tabular}{p{0.8cm}<{\centering}|p{1.3cm}<{\centering}|p{0.8cm}<{\centering}|p{0.8cm}<{\centering}|p{1.3cm}<{\centering}|p{0.8cm}<{\centering}}
\hline
$\mu_{+,1}$&$\mu_{+,2}$&$\mu_{+,3}$&$\mu_{-,1}$&$\mu_{-,2}$&$\mu_{-,3}$\\
\hline
$D_{4}$&$\mathrm{Im}z<0$&$D_{1}$&$D_{3}$&$\mathrm{Im}z>0$&$D_{2}$\\
\hline
\end{tabular}
\end{table}
And (\ref{muphi}) indicates that the same analyticity and boundedness properties also hold for the columns of $\Phi_{\pm}(x,t,z)$.
\end{thm}
\textbf{Proof.}\ \
For a matrix $M$  with the following product notation
\begin{equation}
e^{i(x-y)J(z)}Me^{-i(x-y)J(z)}=\left(
\begin{array}{ccc}
m_{11}&e^{i(k-\lambda)(x-y)}m_{12}&e^{-2i\lambda(x-y)}m_{13}\\
e^{-i(k-\lambda)(x-y)}m_{21}&m_{22}&e^{-i(k+\lambda)(x-y)}m_{23}\\
e^{2i\lambda(x-y)}m_{31}&e^{i(k+\lambda)(x-y)}m_{32}&m_{33}
\end{array}
\right)
\end{equation}
We consider the analytic of the first column.
\begin{equation}
e^{-i(k-\lambda)(x-y)}=e^{\frac{q_{0}^{2}}{|z|^{2}}\mathrm{Im}{z}(x-y)}e^{-i\frac{q_{0}^{2}}{|z|^{2}}\mathrm{Re}{z}(x-y)}
\end{equation}
Then we derived that
\begin{subequations}
\begin{align}
&\mu_{-,1}\quad\mathrm{analytic},\quad\quad \{\mathrm{Im}{z}<0\cup x-y>0\},\\
&\mu_{+,1}\quad\mathrm{analytic},\quad\quad \{\mathrm{Im}{z}>0\cup x-y<0\}.
\end{align}
\end{subequations}
\begin{equation}
e^{2i\lambda(x-y)}=e^{-\frac{|z|^{2}-q_{0}^{2}}{|z|^{2}}\mathrm{Im}{z}(x-y)}e^{2i\mathrm{Re}{\lambda}(x-y)}.
\end{equation}
So we have the following results
\begin{subequations}
\begin{align}
&\mu_{-,1}\quad\mathrm{analytic in} D_{3},\quad\quad \{\mathrm{Im}{z}<0\cup x-y>0\cup|z|^{2}-q_{0}^{2}<0\},\\
&\mu_{+,1}\quad\mathrm{analytic in} D_{4},\quad\quad \{\mathrm{Im}{z}>0\cup x-y<0\cup|z|^{2}-q_{0}^{2}<0\}.
\end{align}
\end{subequations}
The analytic of the second and third column can be derived in the similar way. $\Box$

Now we introduce the scattering matrix $A(z)$. We note that $\mathrm{tr}X=-ik$ and $\mathrm{tr}T=\mathrm{-4ik^{3}}$. If $\Phi(x,t,z)$ is a solution of (\ref{laxpair}), by using Abel's theorem,   we have
\begin{equation}
(\mathrm{det}\Phi)_{x} =\mathrm{tr}X\mathrm{det}\Phi, \quad  (\mathrm{det}\Phi)_{t} =\mathrm{tr}T\mathrm{det}\Phi,\nonumber
\end{equation}
which can be written into  the following differential equations
\begin{equation}
\left[e^{-i\Theta}\mathrm{det}\Phi\right]_{x}=\left[e^{-i\Theta}\mathrm{det}\Phi\right]_{t}=0.\nonumber
\end{equation}
Again by using the  boundary condition (\ref{phigamma}), we  find that
\begin{equation}
\mathrm{det}\Phi_{\pm}(x,t,z)=\gamma(z)e^{i\theta_{2}(x,t,z)}.
\end{equation}
$\Phi_{+}(x,t,z)$ and $\Phi_{-}(x,t,z)$ are the fundamental matrix solutions of the Lax pair. It is indicates the existence of a $3\times3$ matrix $A(z)$, so that
\begin{equation}
\Phi_{-}(x,t,z)=\Phi_{+}(x,t,z)A(z),\label{phia}
\end{equation}
where $A(z)=\left(a_{ij}(z)\right)$ and $\mathrm{det}A(z)=1$. And we introduce the inverse matrix $A^{-1}(z)=B(z)=\left(b_{ij}(z)\right)$.
In the $2\times2$ spectral problem, the analyticity of scattering matrix $A(z)$ can be derived from their representations as Wronskians of (\ref{phia}). In the $3\times3$ spectral problem, however, this approach is no longer applicable. The reason is that the second column of Jost eigenfunction is not analytic in one of analytic region $D_{j}$.
\begin{thm}\label{thm3}
Under the same hypotheses as in \ref{thm1}, the diagonal scattering coefficient can be analytically extended of $\Sigma$ in the following regions:
\renewcommand\arraystretch{2}
\begin{table}[H]
\caption{the analytically of $a_{jj}$ and $b_{jj}$}
\centering
\begin{tabular}{p{0.8cm}<{\centering}|p{1.3cm}<{\centering}|p{0.8cm}<{\centering}|p{0.8cm}<{\centering}|p{1.3cm}<{\centering}|p{0.8cm}<{\centering}}
\hline
$a_{11}$&$a_{22}$&$a_{33}$&$b_{11}$&$b_{22}$&$b_{33}$\\
\hline
$D_{3}$&$\mathrm{Im}z>0$&$D_{2}$&$D_{4}$&$\mathrm{Im}z<0$&$D_{1}$\\
\hline
\end{tabular}
\end{table}
\end{thm}

 \noindent\textbf{Proof.}\ \
We note that the first part of the (\ref{lax}) is equivalent the following problem£º
\begin{equation}
\Phi_{\pm,x}(x,z)=\check{X}_{\pm}\Phi_{\pm}(x,z)+(Q(x)-Q_{f}(x))\Phi_{\pm},\label{checkphi}
\end{equation}
where
\begin{equation}
\check{X}(x,z)=H(x)X_{+}(z)+H(-x)X_{-}(z), \quad Q_{f}(x)=H(x)Q_{+}+H(-x)Q_{-},
\end{equation}
and $H(x)$ is the Heaviside function
\begin{equation}
H(x)=\left\{
\begin{aligned}
&1, &x\geq0\\
&0, & x<0
\end{aligned}\right.\nonumber
\end{equation}
We introduce the fundamental eigenfunction $\check{\Phi}_{\pm}(x,z)$ as the square matrix solution of (\ref{checkphi})
\begin{equation}
\check{\Phi}_{\pm}(x,z)\sim e^{xX_{\pm}(z)},\quad\quad x\rightarrow\pm\infty.
\end{equation}
By solving (\ref{checkphi}) in the similar way as (\ref{muphi}), we introduce the following transformation
\begin{equation}
\check{\mu}_{\pm}=\check{\Phi}_{\pm}e^{-x\check{X}},\nonumber
\end{equation}
and we obtain the full derivative
\begin{equation}
\mathrm{d}(e^{-x\check{X}}\check{\mu}_{\pm}e^{x\check{X}})=e^{-x\check{X}}(Q-Q_{f})\check{\mu}_{\pm}e^{x\check{X}}.\nonumber
\end{equation}
We deriving the integral equation
\begin{subequations}
\begin{align}
&\check{\Phi}_{+}(x,z)=G_{f}(x,0,z)-\int_{x}^{+\infty}G_{f}(x,y,z)[Q(y)-Q_{f}(y)]\check{\Phi}_{+}(y)\mathrm{d}y,\label{phig1}\\
&\check{\Phi}_{-}(x,z)=G_{f}(x,0,z)+\int_{-\infty}^{x}G_{f}(x,y,z)[Q(y)-Q_{f}(y)]\check{\Phi}_{-}(y)\mathrm{d}y,\label{phig2}
\end{align}\label{phig}
\end{subequations}
where
\begin{equation}
G_{f}(x,y,z)=\left\{
\begin{aligned}
&e^{(x-y)X_{+}},& &x,y\geq0\\
&e^{(x-y)X_{-}},& &x,y\leq0\\
&e^{xX_{+}}e^{-yX_{-}},& &x,-y\geq0\\
&e^{xX_{-}}e^{-yX_{+}},& &x,-y\leq0
\end{aligned}\right.\nonumber
\end{equation}
For equation (\ref{phig1}),
\begin{equation}
\check{\Phi}_{\pm}(x,z)=G_{f}(x,0,z)A_{\mp},\label{checkphi}
\end{equation}
where
\begin{equation}
A_{\mp}=I\mp\int_{-\infty}^{+\infty}G_{f}(0,y,z)[Q(y)-Q_{f}(y)]\check{\Phi}_{\pm}\mathrm{d}y.
\end{equation}
Considering the boundary conditions as $x\rightarrow\pm\infty$
\begin{equation}
\check{\Phi}_{\pm}\sim e^{xX_{\pm}}=\Gamma_{\pm}e^{i\Theta x}\Gamma_{\pm}^{-1},\quad
\Phi_{\pm}\sim\Gamma_{\pm}e^{i\Theta}.
\end{equation}
Then we obtain
\begin{equation}
\Phi_{\pm}(x,z)\sim\check{\Phi}_{\pm}(x,z)\Gamma_{\pm}(z),
\end{equation}
so (\ref{phig}) implied that
\begin{subequations}
\begin{align}
&\Phi_{+}(x,z)=G_{f}(x,0,z)\Gamma_{+}(z)-\int_{x}^{+\infty}G_{f}(x,y,z)[Q(y)-Q_{f}(y)]\Phi_{+}(y)\mathrm{d}y,\label{phigamma1}\\
&\Phi_{-}(x,z)=G_{f}(x,0,z)\Gamma_{-}(z)+\int_{-\infty}^{x}G_{f}(x,y,z)[Q(y)-Q_{f}(y)]\Phi_{-}(y)\mathrm{d}y,\label{phigamma2}
\end{align}\label{phigamma12}
\end{subequations}
We compare the asymptotics as $x\rightarrow\infty$ of $\Phi_{-}$ from (\ref{checkphi}) with those of $\Phi_{+}A(z)$ from (\ref{phigamma}) to obtain
\begin{equation}
A(z)=\Gamma_{+}^{-1}(z)A_{+}(z)\Gamma_{-}(z).\label{agamma}
\end{equation}
The equation (\ref{agamma}) implied the following expression of scattering matrix
\begin{equation}
A(z)=\int_{0}^{+\infty}\mathrm{I}\mathrm{d}y+\int_{-\infty}^{0}\mathrm{II}\mathrm{d}y+\Gamma_{+}^{-1}(z)\Gamma_{-}(z)\nonumber
\end{equation}
where the integrands are
\begin{subequations}
\begin{align}
&\mathrm{I}=e^{-iyJ(z)}\Gamma_{+}^{-1}(z)[Q(y)-Q_{+}(y)]\Phi_{-}(y,z),\nonumber\\
&\mathrm{II}=\Gamma_{+}^{-1}(z)\Gamma_{-}(z)e^{-iyJ(z)}\Gamma_{-}^{-1}(z)[Q(y)-Q_{-}(z)]\Phi_{-}(y,z).\nonumber
\end{align}
\end{subequations}
The diagonal elements of $\mathrm{I}$ and $\mathrm{II}$ are, respectively
\begin{subequations}
\begin{align}
&\mathrm{I}_{11}=\sum_{j=1}^{3}\varrho_{1j}e^{iy\lambda}\Phi_{-,j1},\quad
\mathrm{I}_{22}=\sum_{j=1}^{3}\varrho_{2j}e^{iyk}\Phi_{-,j2},\quad
\mathrm{I}_{33}=\sum_{j=1}^{3}\varrho_{3j}e^{-iy\lambda}\Phi_{-,j3}\nonumber\\
&\mathrm{II}_{11}=\sum_{j=1}^{3}\left(\nu_{11}\vartheta_{1j}+\nu_{12}\vartheta_{2j}e^{iy(k-\lambda)}+\nu_{13}\vartheta_{3j}e^{-2iy\lambda}\right)e^{iy\lambda}\Phi_{-,j1}\nonumber\\
&\mathrm{II}_{22}=\sum_{j=1}^{3}\left(\nu_{21}\vartheta_{1j}+\nu_{22}\vartheta_{2j}e^{iy(k-\lambda)}+\nu_{23}\vartheta_{3j}e^{-2iy\lambda}\right)e^{iy\lambda}\Phi_{-,j2}\nonumber\\
&\mathrm{II}_{33}=\sum_{j=1}^{3}\left(\nu_{31}\vartheta_{1j}+\nu_{32}\vartheta_{2j}e^{iy(k-\lambda)}+\nu_{33}\vartheta_{3j}e^{-2iy\lambda}\right)e^{iy\lambda}\Phi_{-,j3}\nonumber
\end{align}
\end{subequations}
where
\begin{equation}
\Gamma_{+}^{-1}\Gamma_{-}=(\nu_{ij}), \quad \Gamma_{-}^{-1}[Q-Q_{-}]=(\vartheta_{ij}), \quad \Gamma_{+}^{-1}[Q-Q_{+}]=(\varrho_{ij}).\nonumber
\end{equation}
$\Box$

\subsection{Auxiliary Eigenfunctions}
The second columns of Jost solution are not analytic in a given domain. To circumvent this defect of analyticity, we introduce a modified Lax pair.
\begin{equation}
\widetilde{\Phi}_{x}=X\widetilde{\Phi},\ \  \widetilde{\Phi}_{t}=T\widetilde{\Phi},\label{modlaxpair}
\end{equation}
where
$\widetilde{X}=\bar{X}(x,t,\bar{z}), \widetilde{T}=\bar{T}(x,t,\bar{z})$. And we note that
\begin{subequations}
\begin{align}
&\bar{Q}=-Q^{T}, \quad Q=-Q^{\dag},\nonumber\\
&Q\Lambda=-\Lambda Q, \quad Q^{T}\Lambda=-\Lambda Q^{T}.\nonumber
\end{align}
\end{subequations}
Before reconstruct a solution of Lax pair (\ref{laxpair}) from the modified Lax pair (\ref{modlaxpair}), we need introduce the following lemma.
\begin{lemma}\label{lemma1}
For $\forall\mathbf{u}, \mathbf{v}\in\mathbb{C}^{3}$ satisfying the following equations:
\begin{subequations}
\begin{align}
&[(\Lambda \mathbf{u})\times \mathbf{v}]+[\mathbf{u}\times(\Lambda \mathbf{v})]-[\mathbf{u}\times \mathbf{v}]-[(\Lambda \mathbf{u})\times(\Lambda \mathbf{v})]=\mathbf{0},\nonumber\\
&\Lambda[\mathbf{u}\times \mathbf{v}]=-[(\Lambda \mathbf{u})\times(\Lambda \mathbf{v})],\nonumber\\
&Q[\mathbf{u}\times \mathbf{v}]+[(Q^{T}\mathbf{u})\times \mathbf{v}]+[\mathbf{u}\times(Q^{T}\mathbf{v})]=\mathbf{0},\nonumber\\
&\Lambda Q^{2}[\mathbf{u}\times \mathbf{v}]+[(\Lambda(Q^{T})^{2})\times \mathbf{v}]+[\mathbf{u}\times(\Lambda(Q^{T})^{2})\mathbf{v}]=\mathbf{0},\nonumber
\end{align}
\end{subequations}
where "$\times$" denotes the usual cross product.
\end{lemma}
Using above identities and Lemma\ref{lemma1} one can straightforward to prove the following theorem.
\newtheorem{theorem}{Theorem}
\begin{thm}\label{thm2}
If $\widetilde{\mathbf{v}}(x,t,z)$ and $\widetilde{\mathbf{w}}(x,t,z)$ are the two arbitrary solutions of modified Lax pair (\ref{modlaxpair}), then
\begin{equation}
\mathbf{u}(x,t,z)=e^{i\theta_{2}(x,t,z)}[\widetilde{\mathbf{v}}\times\widetilde{\mathbf{w}}](x,t,z)\nonumber
\end{equation}
is also a solution of Lax pair (\ref{laxpair}).
\end{thm}

\noindent\textbf{Proof.}\ \
Through some calculations, we have
\begin{equation}
\begin{split}
[\widetilde{\mathbf{v}}\times\widetilde{\mathbf{w}}]_{x}=&\widetilde{\mathbf{v}}_{x}\times\widetilde{\mathbf{w}}+\widetilde{\mathbf{v}}\times\widetilde{\mathbf{w}}_{x}\\
=&(ik\Lambda+\bar{Q})\widetilde{\mathbf{v}}\times\widetilde{\mathbf{w}}+\widetilde{\mathbf{v}}\times(ik\Lambda+\bar{Q})\widetilde{\mathbf{w}}\\
=&ik(\Lambda\widetilde{\mathbf{v}})\times\widetilde{\mathbf{w}}+(\bar{Q}\widetilde{\mathbf{v}})\times\widetilde{\mathbf{w}}
+ik[\widetilde{\mathbf{v}}\times(\Lambda\widetilde{\mathbf{w}})]+[\widetilde{\mathbf{v}}\times(\bar{Q}\widetilde{\mathbf{w}})]\\
=&ik\{[(\Lambda\widetilde{\mathbf{v}})\times\widetilde{\mathbf{w}}]+[\widetilde{\mathbf{v}}\times(\Lambda\widetilde{\mathbf{w}})]\}
+\{[(\bar{Q}\widetilde{\mathbf{v}})\times\widetilde{\mathbf{w}}]+[\widetilde{\mathbf{v}}\times(\bar{Q}\widetilde{\mathbf{w}})]\},
\end{split}\nonumber
\end{equation}
And we note that
\begin{equation}
[(\bar{Q}\widetilde{\mathbf{v}})\times\widetilde{\mathbf{w}}]+[\widetilde{\mathbf{v}}\times(\bar{Q}\widetilde{\mathbf{w}})]=Q[\widetilde{\mathbf{v}}\times\widetilde{\mathbf{w}}].\nonumber
\end{equation}
So we derived the $x$-part
\begin{eqnarray}
\begin{split}
\mathbf{u}_{x}(x,t,z)&=e^{i\theta_{2}}(i\theta_{2,x})[\widetilde{\mathbf{v}}\times\widetilde{\mathbf{w}}]+e^{i\theta_{2}}[\widetilde{\mathbf{v}}\times\widetilde{\mathbf{w}}]_{x}\\
&=e^{i\theta_{2}}(-ik)[\widetilde{\mathbf{v}}\times\widetilde{\mathbf{w}}]+e^{i\theta_{2}}(ik[\widetilde{\mathbf{v}}\times\widetilde{\mathbf{w}}]+ik[(\Lambda\widetilde{\mathbf{v}})\times(\Lambda\widetilde{\mathbf{w}})])+e^{i\theta_{2}}Q[\widetilde{\mathbf{v}}\times\widetilde{\mathbf{w}}]\\
&=-ik\Lambda e^{i\theta_{2}}[\widetilde{\mathbf{v}}\times\widetilde{\mathbf{w}}]+e^{i\theta_{2}}Q[\widetilde{\mathbf{v}}\times\widetilde{\mathbf{w}}]\\
&=(-ik\Lambda+Q)e^{i\theta_{2}}[\widetilde{\mathbf{v}}\times\widetilde{\mathbf{w}}].\nonumber
\end{split}
\end{eqnarray}
In the similar way, we can derived the $t$-part. $\Box$

As $x\rightarrow\infty$, we obtain the asymptotic spectral problem of (\ref{modlaxpair})
\begin{equation}
\widetilde{\Phi}_{\pm,x}=\widetilde{X}_{\pm}\widetilde{\Phi}_{\pm},\quad \widetilde{\Phi}_{\pm,t}=\widetilde{T}_{\pm}\widetilde{\Phi}_{\pm}.\nonumber
\end{equation}
The eigenvalues of $\widetilde{X}_{\pm}$ and $\widetilde{T}_{\pm}$ are
\[\begin{array}{ccc}
ik, & -i\lambda, & i\lambda\\
4ik^{3}, & -2i\lambda(2k^{2}-q_{0}^{2}), & 2i\lambda(2k^{2}-q_{0}^{2})
\end{array}\]
We can deriving the eigenvector $\widetilde{\Gamma}_{\pm}(z)=\bar{\Gamma}_{\pm}(\bar{z})$ of matrix $\widetilde{X}_{\pm}$ and $\widetilde{T}_{\pm}$. And we note that $\mathrm{det}\widetilde{\Gamma}_{\pm}(z)=\gamma(z)$.
For $z\in\Sigma$, we define the Jost solutions of (\ref{modlaxpair}) as the simultaneous solutions such that
\begin{equation}
\widetilde{\Phi}_{\pm}(x,t,z)\sim\widetilde{\Gamma}_{\pm}(z)e^{-i\Theta(x,t,z)},\quad x\rightarrow\infty,\quad z\in\Sigma.\nonumber
\end{equation}
As in section 2.2, we introducing the modified adjoint eigenfunction
\begin{equation}
\widetilde{\mu}_{\pm}(x,t,z)=\widetilde{\Phi}_{\pm}(x,t,z)e^{i\Theta(x,t,z)}.
\end{equation}
One can show the analyticity of $\widetilde\mu_{\pm}$
\renewcommand\arraystretch{2}
\begin{table}[H]
\caption{the analytically of $\widetilde{\mu}_{\pm}$}
\centering
\begin{tabular}{p{0.8cm}<{\centering}|p{1.3cm}<{\centering}|p{0.8cm}<{\centering}|p{0.8cm}<{\centering}|p{1.3cm}<{\centering}|p{0.8cm}<{\centering}}
\hline
$\widetilde{\mu}_{+,1}$&$\widetilde{\mu}_{+,2}$&$\widetilde{\mu}_{+,3}$&$\widetilde{\mu}_{-,1}$&$\widetilde{\mu}_{-,2}$&$\widetilde{\mu}_{-,3}$\\
\hline
$D_{3}$&$\mathrm{Im}z>0$&$D_{2}$&$D_{4}$&$\mathrm{Im}z<0$&$D_{1}$\\
\hline
\end{tabular}
\end{table}
$\widetilde{\Phi}_{\pm}(x,t,z)$ are both the fundamental matrix solutions of the same problem, so we introduce the adjoint scattering matrix as
\begin{equation}
\widetilde{\Phi}_{-}(x,t,z)=\widetilde{\Phi}_{+}(x,t,z)\widetilde{A}(z).\label{widephia}
\end{equation}

Under the same hypotheses as in \ref{thm1}, the diagonal scattering coefficient can be analytically extended of $\Sigma$ in the following regions:
\renewcommand\arraystretch{2}
\begin{table}[H]
\caption{the analytically of $\widetilde{a}_{jj}$ and $\widetilde{b}_{jj}$}
\centering
\begin{tabular}{p{0.8cm}<{\centering}|p{1.3cm}<{\centering}|p{0.8cm}<{\centering}|p{0.8cm}<{\centering}|p{1.3cm}<{\centering}|p{0.8cm}<{\centering}}
\hline
$\widetilde{a}_{11}$&$\widetilde{a}_{22}$&$\widetilde{a}_{33}$&$\widetilde{b}_{11}$&$\widetilde{b}_{22}$&$\widetilde{b}_{33}$\\
\hline
$D_{4}$&$\mathrm{Im}z<0$&$D_{1}$&$D_{3}$&$\mathrm{Im}z>0$&$D_{2}$\\
\hline
\end{tabular}
\end{table}

In light of above results, we can define four auxiliary eigenfunctions $\chi_{j}, (j=1,2,3,4)$ as the new solutions of the Lax pair (\ref{laxpair})
\begin{subequations}
\begin{align}
&\chi_{1}(x,t,z)=e^{i\theta_{2}}[\widetilde{\Phi}_{+,2}\times\widetilde{\Phi}_{-,3}](x,t,z),\\
&\chi_{2}(x,t,z)=e^{i\theta_{2}}[\widetilde{\Phi}_{-,2}\times\widetilde{\Phi}_{+,3}](x,t,z),\\
&\chi_{3}(x,t,z)=e^{i\theta_{2}}[\widetilde{\Phi}_{+,1}\times\widetilde{\Phi}_{-,2}](x,t,z),\\
&\chi_{4}(x,t,z)=e^{i\theta_{2}}[\widetilde{\Phi}_{-,1}\times\widetilde{\Phi}_{+,2}](x,t,z).
\end{align}\label{chi}
\end{subequations}
We note that $\chi_{j}$ is analyticity for $z\in D_{j}$. And there are some simple relations exists between the adjoint Jost solutions eigenfunction and the eigenfunctions of the Lax pair (\ref{laxpair}):

\begin{lemma}\label{lemma2}
For $z\in\Sigma$ and for all cyclic indices $j$, $l$, and $m$
\begin{subequations}
\begin{align}
&\Phi_{\pm,j}(x,t,z)=e^{i\theta_{2}}[\widetilde{\Phi}_{\pm,l}\times\widetilde{\Phi}_{\pm,m}](x,t,z)/\gamma_{j}(z),\label{phij}\\
&\widetilde{\Phi}_{\pm,j}(x,t,z)=e^{-i\theta_{2}}[\Phi_{\pm,l}\times\Phi_{\pm,m}](x,t,z)/\gamma_{j}(z),\label{widephij}
\end{align}
\end{subequations}
where $\gamma_{1}(z)=\gamma_{3}(z)=1$, $\gamma_{2}(z)=\gamma(z)$.
\end{lemma}

\noindent\textbf{Proof.}\ \
We prove (\ref{phij}) with $j=3$, the rest of (\ref{lemma2})is verified in the similar method. From (\ref{phigamma}) and $\mathrm{Theorem}\ref{thm2}$, we deriving
\begin{equation}
\mathrm{u}_{\pm}(x,t,z)=e^{-i\theta_{1}(x,t,z)}\Gamma_{\pm,3}(z), \quad\quad x\rightarrow\infty.\nonumber
\end{equation}
The vector function $\mathrm{u}_{\pm}(x,t,z)$ must be linear combination of the columns of $\Phi_{\pm}$, then there exist three scalar functions $a_{\pm}(z)$, $b_{\pm}(z)$ and $c_{\pm}(z)$ so that
\begin{equation}
\mathrm{u}_{\pm}(x,t,z)=a_{\pm}(z)\Phi_{\pm,1}(x,t,z)+b_{\pm}(z)\Phi_{\pm,2}(x,t,z)+c_{\pm}(z)\Phi_{\pm,3}(x,t,z).\nonumber
\end{equation}
Comparing with the asymptotics of $\Phi_{\pm}(x,t,z)$ as $x\rightarrow\pm\infty$ in (\ref{phigamma}) yields $a_{\pm}(z)=b_{\pm}(z)=0$ and $c_{\pm}(z)=1$. $\Box$

It is naturally that these relations in Lemma \ref{lemma2} imply the relation of scattering matrix as in Theorem\ref{thm5}:
\begin{thm}\label{thm5}
The scattering matrix $A(z)$ and $\widetilde{A}(z)$ admit the relation
\begin{equation}
\widetilde{A}(z)=D(z)(A^{-1}(z))^{T}D^{-1}(z).\label{widea}
\end{equation}
\end{thm}

\noindent\textbf{Proof.}\ \
Equation (\ref{phia}) and $\mathrm{Lemma}\ref{lemma2}$ yield
\begin{equation}
\widetilde{\Phi}_{-,1}=(a_{22}a_{33}-a_{32}a_{23})\widetilde{\Phi}_{+,1}+\gamma(a_{32}a_{13}-a_{12}a_{33})\widetilde{\Phi}_{+,2}+(a_{12}a_{23}-a_{22}a_{13})\widetilde{\Phi}_{+,3}.\nonumber
\end{equation}
Comparing with (\ref{widephia}) yields
\begin{equation}
\widetilde{a}_{11}=a_{22}a_{33}-a_{32}a_{23},\quad \widetilde{a}_{21}=\gamma(a_{32}a_{13}-a_{12}a_{33}),\quad \widetilde{a}_{31}=a_{12}a_{23}-a_{22}a_{13}.\nonumber
\end{equation}
In the similar way, we derived
\begin{subequations}
\begin{align}
&\widetilde{a}_{12}=\frac{1}{\gamma}(a_{23}a_{31}-a_{33}a_{21}),\quad\widetilde{a}_{22}=a_{33}a_{11}-a_{13}a_{31},\quad\widetilde{a}_{32}=\frac{1}{\gamma} (a_{13}a_{21}-a_{23}a_{11}),\nonumber\\
&\widetilde{a}_{13}=a_{21}a_{32}-a_{31}a_{22},\quad \widetilde{a}_{23}=\gamma(a_{31}a_{12}-a_{11}a_{32}),\quad \widetilde{a}_{33}=a_{11}a_{22}-a_{21}a_{12}.\nonumber
\end{align}
\end{subequations}
We note that
\begin{equation}
(A^{-1}(z))^{T}=\left(
\begin{array}{ccc}
a_{22}a_{33}-a_{23}a_{32}&a_{23}a_{31}-a_{21}a_{33}&a_{21}a_{32}-a_{22}a_{31}\\
a_{13}a_{32}-a_{12}a_{33}&a_{11}a_{33}-a_{13}a_{31}&a_{12}a_{31}-a_{11}a_{32}\\
a_{12}a_{23}-a_{13}a_{22}&a_{13}a_{21}-a_{11}a_{23}&a_{11}a_{22}-a_{12}a_{21}
\end{array}
\right).
\nonumber\end{equation}
We finally obtain
\begin{equation}
\widetilde{A}(z)=D(z)(A^{-1}(z))^{T}D^{-1}(z).\nonumber
\end{equation}
$\Box$

Next, combining Lemma \ref{lemma2} and the equation (\ref{widephia}) in the definition (\ref{chi}) deriving Theorem\ref{thm6}:
\begin{thm}\label{thm6}
For $z\in\Sigma$, the Jost eigenfunction have the following relations
\begin{subequations}
\begin{align}
&\Phi_{-,1}=\frac{1}{a_{22}}[\chi_{1}+a_{21}\Phi_{-,2}]=\frac{1}{a_{33}}[a_{31}\Phi_{-,3}+\chi_{2}],\\
&\Phi_{-,3}=\frac{1}{a_{22}}[\chi_{4}+a_{23}\Phi_{-,2}]=\frac{1}{a_{11}}[a_{13}\Phi_{-,3}+\chi_{3}],\\
&\Phi_{+,1}=\frac{1}{b_{22}}[\chi_{2}+b_{21}\Phi_{+,2}]=\frac{1}{b_{33}}[b_{31}\Phi_{+,3}+\chi_{1}],\\
&\Phi_{+,3}=\frac{1}{b_{22}}[\chi_{3}+b_{23}\Phi_{+,2}]=\frac{1}{b_{11}}[b_{13}\Phi_{+,1}+\chi_{4}].
\end{align}
\end{subequations}
For convenient, we omit the independent variable.
\end{thm}

\noindent\textbf{Proof.}\ \
Substituting (\ref{widephia}) into (\ref{chi}) obtain the following equations
\begin{subequations}
\begin{align}
&\chi_{1}=\widetilde{b}_{22}e^{i\theta_{2}}[\widetilde{\Phi}_{-,1}\times\widetilde{\Phi}_{-,2}]+\widetilde{b}_{32}e^{i\theta_{2}}[\widetilde{\Phi}_{-,1}\times\widetilde{\Phi}_{-,3}],\\
&\chi_{4}=\widetilde{b}_{12}e^{i\theta_{2}}[\widetilde{\Phi}_{-,1}\times\widetilde{\Phi}_{-,3}]+\widetilde{b}_{22}e^{i\theta_{2}}[\widetilde{\Phi}_{-,2}\times\widetilde{\Phi}_{-,3}],
\end{align}\label{chi14}
\end{subequations}
Substituting (\ref{phij}) into (\ref{chi14}) yields
\begin{subequations}
\begin{align}
\chi_{1}=\widetilde{b}_{22}\Phi_{-,3}-\widetilde{b}_{32}\gamma\Phi_{-,2},\\
\chi_{4}=\widetilde{b}_{12}\Phi_{-,2}+\widetilde{b}_{22}\gamma\Phi_{-,1}.
\end{align}\label{chiphi}
\end{subequations}
Applying (\ref{widea}) to (\ref{chiphi}), we obtain
\begin{equation}
\Phi_{-,1}=\frac{1}{a_{22}}(\chi_{4}+a_{21}\Phi_{-,2}),\quad \Phi_{-,3}=\frac{1}{a_{22}}(\chi_{1}+a_{23}\Phi_{-,2}).\nonumber
\end{equation}
In the similar way we can obtain the rest of (\ref{thm6}).
$\Box$

To remove the exponential oscillations, we define the modified auxiliary eigenfunctions
\begin{subequations}
\begin{align}
&m_{j}(x,t,z)=e^{-i\theta_{1}(x,t,z)}\chi_{j}(x,t,z), \quad j=1,2\\
&m_{j}(x,t,z)=e^{i\theta_{1}(x,t,z)}\chi_{j}(x,t,z), \quad j=3,4
\end{align}
\end{subequations}

\section{Symmetries}
For the zero boundary conditions with the only one  the symmetry $k\mapsto \bar{k}$. However, for the nonzero boundary conditions, the symmetries are complicated because the Riemann surface introduced. It not only involved with the symmetries $k\mapsto \bar{k}$, but also involved with the symmetries $k\mapsto -\frac{q_{0}^{2}}{z}$. It namely that mapping the upper-half plane into the lower-half plane and mapping the exterior of circle $C_{0}$ of radius $q_{0}$ into the interior.
\subsection{The First Symmetries}
In this subsection, we firstly consider the map that $z\mapsto\bar{z}$.

\begin{lemma}\label{lemma3}
If $\Phi(x,t,z)$ is a non-singular solution of the Lax pair (\ref{laxpair}), so is $\mathrm{w}(x,t,z)=(\Phi^{\dag}(x,t,\bar{z}))^{-1}$.
\end{lemma}
We also show that

\noindent  {\rm\bf Proof.}  If $\Phi(x,t,z)$ is a non-singular solution of the Lax pair. Then we have
\begin{equation}
\Phi^{\dag}_{x}=\Phi^{\dag}X^{\dag}, \quad \Phi^{\dag}_{t}=\Phi^{\dag}T^{\dag}.\label{phih}
\end{equation}
If we substituting $X$ and $T$ into (\ref{phih}), we have
\begin{subequations}
\begin{align}
\mathrm{w}_{x}=&-[\Phi^{\dag}(\bar{z})]^{-1}[\Phi^{\dag}(\bar{z})]_{x}[\Phi^{\dag}(\bar{z})]^{-1}\nonumber\\
=&-[\Phi^{\dag}(\bar{z})]^{-1}\Phi^{\dag}(\bar{z})(ik\Lambda-Q)[\Phi^{\dag}(\bar{z})]^{-1}\nonumber\\
=&X\mathrm{w},\nonumber
\end{align}
\end{subequations}
\begin{subequations}
\begin{align}
\mathrm{w}_{t}=&-[\Phi^{\dag}(\bar{z})]^{-1}[\Phi^{\dag}(\bar{z})]_{t}[\Phi^{\dag}(\bar{z})]^{-1}\nonumber\\
=&-[\Phi^{\dag}(\bar{z})]^{-1}\Phi^{\dag}(\bar{z})\left\{4ik^{3}\Lambda-4k^{2}Q-2ik(-Q_{x}-Q^{2})\Lambda-2Q^{3}+Q_{xx}-[Q_{x},Q]\right\}[\Phi^{\dag}(\bar{z})]^{-1}\nonumber\\
=&T\mathrm{w}.\nonumber
\end{align}
\end{subequations}
Then $\mathrm{w}(x,t,z)$ is a solution of the Lax pair. $\Box$

The Jost eigenfunctions satisfy the symmetry
\begin{lemma}\label{lemma4}
\begin{equation}
(\Phi^{\dag}_{\pm}(x,t,\bar{z}))^{-1}C(z)=\Phi_{\pm}(x,t,z),
\end{equation}
where $C(z)=\mathrm{diag}(\gamma(z),1,\gamma(z))$.
\end{lemma}

\noindent\textbf{Proof.}\ \
From the Lemma \ref{lemma3}, we known that $\Phi_{\pm}(x,t,z)$ and $(\Phi_{\pm}^{\dag}(x,t,\bar{z}))^{-1}$ are the solutions of the Lax pair. And, we note that
\begin{equation}
(e^{i\Theta(x,t,\bar{z})})^{\dag}=e^{-i\Theta(x,t,z)}.\nonumber
\end{equation}
\begin{equation}
\Phi_{\pm}(x,t,z)\sim\Gamma_{\pm}e^{i\Theta(z)}, \quad x\rightarrow\pm\infty
\end{equation}
\begin{subequations}
\begin{align}
(\Phi_{\pm}^{\dag}(x,t,\bar{z}))^{-1}\sim&((\Gamma_{\pm}(\bar{z})e^{i\Theta(\bar{z})})^{\dag})^{-1}\nonumber\\
&=(e^{-i\Theta(z)}\Gamma_{\pm}^{\dag}(\bar{z}))^{-1}
=(\Gamma_{\pm}^{\dag}(\bar{z}))^{-1}e^{i\Theta(z)}\nonumber\\
&=(C(z)\Gamma_{\pm}^{-1}(z))^{-1}e^{i\Theta(z)}
= \Gamma_{\pm}(z)C^{-1}(z)e^{i\Theta(z)}\nonumber\\
=&\Gamma_{\pm}(z)e^{i\Theta(z)}C^{-1}(z)\nonumber
\end{align}
\end{subequations}
Then, we have
\begin{equation}
(\Phi^{\dag}_{\pm}(x,t,\bar{z}))^{-1}C(z)\sim\Phi_{\pm}(x,t,z), \quad x\rightarrow\pm\infty.\nonumber
\end{equation}
$\Box$

We also can show  that
\begin{lemma}\label{lemma19}
For the columns of matrix $\Phi_{\pm}(z,t,z)$ with the following property about cross product
\begin{equation}
(\Phi^{-1}_{\pm}(x,t,z))^{T}=\frac{1}{\det\Phi_{\pm}(x,t,z)}(\Phi_{\pm,2}\times\Phi_{\pm,3},\Phi_{\pm,3}\times\Phi_{\pm,1},\Phi_{\pm,1}\times\Phi_{\pm,2})(x,t,z).\label{phit}
\end{equation}
\end{lemma}
Using the Theorem \ref{thm6} and Lemma \ref{lemma4} yields:
\begin{lemma}\label{lemma5}
The Jost eigenfunctions obey the symmetry relations:
\begin{subequations}
\begin{align}
&\bar{\Phi}_{+,1}(x,t,\bar{z})=\frac{e^{-i\theta_{2}}}{b_{22}}[\Phi_{+,2}\times\chi_{3}](x,t,z),\\
&\bar{\Phi}_{-,1}(x,t,\bar{z})=\frac{e^{-i\theta_{2}}}{a_{22}}[\Phi_{-,2}\times\chi_{4}](x,t,z),\\
&\bar{\Phi}_{+,2}(x,t,\bar{z})=\frac{e^{-i\theta_{2}}}{\gamma b_{11}}[\chi_{4}\times\Phi_{+,1}](x,t,z)=\frac{e^{-i\theta_{2}}}{\gamma b_{33}}[\Phi_{+,3}\times\chi_{1}](x,t,z),\\
&\bar{\Phi}_{-,2}(x,t,\bar{z})=\frac{e^{-i\theta_{2}}}{\gamma a_{11}}[\chi_{3}\times\Phi_{-,1}](x,t,z)=\frac{e^{-i\theta_{2}}}{\gamma a_{33}}[\Phi_{-,3}\times\chi_{2}](x,t,z),\\
&\bar{\Phi}_{+,3}(x,t,\bar{z})=\frac{e^{-i\theta_{2}}}{b_{22}}[\chi_{2}\times\Phi_{+,2}](x,t,z),\\
&\bar{\Phi}_{-,3}(x,t,\bar{z})=\frac{e^{-i\theta_{2}}}{a_{22}}[\chi_{1}\times\Phi_{-,2}](x,t,z).
\end{align}\label{barphi}
\end{subequations}
\end{lemma}
Using the equation (\ref{phia}) and Lemma \ref{lemma4} yields:
\begin{lemma}\label{lemma6}
The scattering matrix and its inverse satisfy the symmetry relation
\begin{equation}
A^{\dag}(\bar{z})=C(z)B(z)C^{-1}(z), \quad\quad z\in\Sigma.
\end{equation}
\end{lemma}
For all $z\in\Sigma$, the componentwise satisfy:
\begin{subequations}
\begin{align}
&b_{11}(z)=\bar{a}_{11}(\bar{z}), \quad\quad\quad b_{12}(z)=\frac{\bar{a}_{21}(\bar{z})}{\gamma(z)}, \quad\quad b_{13}(z)=\bar{a}_{31}(\bar{z}),\nonumber\\
&b_{21}(z)=\gamma(z)\bar{a}_{12}(\bar{z}),\quad b_{22}(z)=\bar{a}_{22}(\bar{z}),\quad\quad b_{23}(z)=\gamma(z)\bar{a}_{32}(\bar{z}),\nonumber\\
&b_{31}(z)=\bar{a}_{13}(\bar{z}), \quad\quad\quad b_{32}(z)=\frac{\bar{a}_{23}(\bar{z})}{\gamma(z)}, \quad\quad b_{33}(z)=\bar{a}_{33}(\bar{z}).\nonumber
\end{align}\nonumber
\end{subequations}
And we also derived the symmetry relations of the auxiliary eigenfunctions:
\begin{lemma}\label{lemma7}
The auxiliary eigenfunctions holds the symmetry relations:
\begin{subequations}
\begin{align}
&\bar{\chi}_{1}(x,t,\bar{z})=e^{-i\theta_{2}(z)}[\Phi_{+,2}\times\Phi_{-,3}](x,t,z),\\
&\bar{\chi}_{2}(x,t,\bar{z})=e^{-i\theta_{2}(z)}[\Phi_{-,2}\times\Phi_{+,3}](x,t,z),\\
&\bar{\chi}_{3}(x,t,\bar{z})=e^{-i\theta_{2}(z)}[\Phi_{+,1}\times\Phi_{-,2}](x,t,z),\\
&\bar{\chi}_{4}(x,t,\bar{z})=e^{-i\theta_{2}(z)}[\Phi_{-,1}\times\Phi_{+,2}](x,t,z).
\end{align}
\end{subequations}
\end{lemma}
The proof of Lemma \ref{lemma7} yields:
\begin{equation}
\bar{\Phi}_{\pm,j}(x,t,\bar{z})=e^{-i\theta_{2}(x,t,z)}[\Phi_{\pm,l}\times\Phi_{\pm,m}](x,t,z)/\gamma_{j},
\end{equation}
where $j$, $l$, and $m$ are cyclic indices and $z\in\Sigma$.
\subsection{The Second Symmetries}
In this subsection, we consider the map $z\mapsto -\frac{q_{0}^{2}}{z}$.
\begin{lemma}\label{lemma8}
For all $z\in\Sigma$, the Jost eigenfunctions holds the symmetry relations
\begin{equation}
\Phi_{\pm}(x,t,z)=\Phi_{\pm}(x,t,-\frac{q_{0}^{2}}{z})\Pi(z),
\end{equation}
where
\begin{equation}
\Pi(z)=\left(
\begin{array}{ccc}
0&0&\frac{iq_{0}}{z}\\
0&1&0\\
\frac{iq_{0}}{z}&0&0
\end{array}
\right).
\end{equation}
\end{lemma}
So we derived all of the above relations:
\begin{subequations}
\begin{align}
&\Phi_{\pm,1}(x,t,z)=\frac{iq_{0}}{z}\Phi_{\pm,3}(x,t,-\frac{q_{0}^{2}}{z}),\\
&\Phi_{\pm,2}(x,t,z)=\Phi_{\pm,2}(x,t,-\frac{q_{0}^{2}}{z}),\\
&\Phi_{\pm,3}(x,t,z)=\frac{iq_{0}}{z}\Phi_{\pm,1}(x,t,-\frac{q_{0}^{2}}{z}).
\end{align}\label{phi123}
\end{subequations}
\begin{lemma}\label{9}
The scattering matrix holds the symmetry
\begin{equation}
A(-\frac{q_{0}^{2}}{z})=\Pi(z)A(z)\Pi^{-1}(z),\quad B(-\frac{q_{0}^{2}}{z})=\Pi(z)B(z)\Pi^{-1}(z).
\end{equation}
\end{lemma}
We have the following componentwise relations
\begin{subequations}
\begin{align}
&a_{11}(z)=a_{33}(-\frac{q_{0}^{2}}{z}),\quad a_{12}(z)=\frac{z}{iq_{0}}a_{32}(-\frac{q_{0}^{2}}{z}),a_{13}(z)=a_{31}(-\frac{q_{0}^{2}}{z}),\nonumber\\
&a_{21}(z)=\frac{iq_{0}}{z}a_{23}(-\frac{q_{0}^{2}}{z}), a_{22}(z)=a_{22}(-\frac{q_{0}^{2}}{z}),\quad a_{23}(z)=\frac{iq_{0}}{z}a_{21}(-\frac{q_{0}^{2}}{z}),\nonumber\\
&a_{31}(z)=a_{13}(-\frac{q_{0}^{2}}{z}),\quad a_{32}(z)=\frac{z}{iq_{0}}a_{12}(-\frac{q_{0}^{2}}{z}),a_{33}(z)=a_{11}(-\frac{q_{0}^{2}}{z}).\nonumber
\end{align}
\end{subequations}

\begin{lemma}\label{lemma9}
The auxiliary eigenfunctions holds the symmetries relations
\begin{subequations}
\begin{align}
&\chi_{1}(z)=\frac{iq_{0}}{z}\chi_{4}(-\frac{q_{0}^{2}}{z}),\quad z\in D_{1}\label{symchi14}\\
&\chi_{2}(z)=\frac{iq_{0}}{z}\chi_{3}(-\frac{q_{0}^{2}}{z}),\quad z\in D_{2}.\label{symchi23}
\end{align}
\end{subequations}
\end{lemma}

\subsection{The Third Symmetry}
In this subsection, we introduce the symmetries of reflection coefficients.
\begin{subequations}
\begin{align}
&\rho_{1}(z)=\frac{a_{21}(z)}{a_{11}(z)}=\frac{\gamma(z)\bar{b}_{12}(\bar{z})}{\bar{b}_{11}(\bar{z})},\quad\rho_{2}(z)=\frac{a_{31}(z)}{a_{11}(z)}=\frac{\bar{b}_{13}(\bar{z})}{\bar{b}_{11}(\bar{z})},\\
&\rho_{3}(z)=\frac{a_{32}(z)}{a_{22}(z)}=\frac{\bar{b}_{23}(\bar{z})}{\gamma(z)\bar{b}_{22}(\bar{z})},\quad\rho_{1}(-\frac{q_{0}^{2}}{z})=\frac{z}{iq_{0}}\frac{a_{23}(z)}{a_{33}(z)}=\frac{z}{iq_{0}}\frac{\gamma(z)\bar{b}_{32}(\bar{z})}{\bar{b}_{33}(\bar{z})},\\
&\rho_{2}(-\frac{q_{0}^{2}}{z})=\frac{a_{13}(z)}{a_{33}(z)}=\frac{\bar{b}_{31}(\bar{z})}{\bar{b}_{33}(\bar{z})},\quad\rho_{3}(-\frac{q_{0}^{2}}{z})=\frac{iq_{0}}{z}\frac{a_{12}(z)}{a_{22}(z)}=\frac{iq_{0}}{z}\frac{\bar{b}_{21}(\bar{z})}{\gamma(z)\bar{b}_{22}(\bar{z})},
\end{align}\label{rho}
\end{subequations}

\section{Asymptotic behavior as $z\rightarrow\infty$ and $z\rightarrow 0$}
In order to normalize the Riemann-Hilbert problem, it is necessary to consider the asymptotic behavior of the eigenfunctions and scattering coefficients as $k\rightarrow\infty$, which   corresponds to  the behavior both $z\rightarrow\infty$ and $z\rightarrow 0$. So we consider the following expansion for $\Gamma^{-1}_{\pm}\mu_{\pm}$
\begin{subequations}
\begin{align}
&\Gamma_{\pm}^{-1}\mu_{\pm}=E_{0}+\frac{E_{1}}{z}+\frac{E_{2}}{z^{2}}+\cdots \quad z\rightarrow\infty,\label{expaninfty}\\
&\Gamma_{\pm}^{-1}\mu_{\pm}=F_{0}+F_{1}z+F_{2}z^{2}+\cdots \quad z\rightarrow 0,\label{expan0}
\end{align}\label{expan}
\end{subequations}
where $E_{j}$ and $F_{j}$ are $3\times 3$ matrices and independent of $z$. Through some explicitly evaluate, we obtain the asymptotic behavior:
\begin{thm}\label{thm7}
The asymptotic of $\mu_{\pm}$ as $z\rightarrow\infty$ and $z\rightarrow 0$
\begin{subequations}
\begin{align}
&\mu_{\pm}(x,t,z)=\left(
\begin{array}{ccc}
-\frac{\mathbf{q}_{\pm}}{q_{0}}&\frac{\mathbf{q}_{\pm}^{\bot}}{q_{0}}&-\frac{i\mathbf{q}}{z}\\
\frac{i\mathbf{q}_{\pm}^{\dag}\mathbf{q}}{zq_{0}}&\frac{i(\mathbf{q}_{\pm}^{\bot})^{\dag}\mathbf{q}}{zq_{0}}&1
\end{array}
\right)+\frac{1}{z}[\Gamma_{\pm}]_{bd}[E_{1}]_{bd}+o(\frac{1}{z^{2}}),\quad z\rightarrow\infty.\\
&\mu_{\pm}(x,t,z)=\left(
\begin{array}{ccc}
-\frac{\mathbf{q}}{q_{0}}&\frac{\mathbf{q}_{\pm}^{\bot}}{q_{0}}&-\frac{i\mathbf{q}_{\pm}}{z}\\
\frac{iq_{0}}{z}&0&\frac{\mathbf{q}_{\pm}^{\dag}\mathbf{q}}{q_{0}^{2}}
\end{array}
\right)+[\Gamma_{\pm}]_{bd}[F_{0}]_{bo}+o(1), \quad z\rightarrow 0.
\end{align}
\end{subequations}
\end{thm}

\begin{thm}\label{thm8}
The asymptotic of $m_{j}$ as $z\rightarrow\infty$
\begin{subequations}
\begin{align}
&m_{1}(x,t,z)=\left(
\begin{array}{c}
-\frac{\mathbf{q}_{+}}{q_{0}}\\
\frac{i\mathbf{q}_{+}^{\dag}\mathbf{q}}{zq_{0}}
\end{array}
\right)+o(\frac{1}{z^{2}}), \quad m_{2}(x,t,z)=\left(
\begin{array}{c}
-\frac{\mathbf{q}_{-}}{q_{0}}\\
\frac{i\mathbf{q}_{-}^{\dag}\mathbf{q}}{zq_{0}}
\end{array}
\right)+o(\frac{1}{z^{2}}),\\
&m_{3}(x,t,z)=\left(
\begin{array}{c}
-\frac{i\mathbf{q}}{zq_{0}^{2}}[\mathbf{q}_{-}^{\dag}\mathbf{q}_{+}]\\
\frac{1}{q_{0}^{2}}[\mathbf{q}_{-}^{\dag}\mathbf{q}_{+}]
\end{array}
\right)+o(\frac{1}{z^{2}}), \quad m_{4}(x,t,z)=\left(
\begin{array}{c}
-\frac{i\mathbf{q}}{zq_{0}^{2}}[\mathbf{q}_{-}^{\dag}\mathbf{q}_{+}]\\
\frac{1}{q_{0}^{2}}[\mathbf{q}_{-}^{\dag}\mathbf{q}_{+}]
\end{array}
\right)+o(\frac{1}{z^{2}})
\end{align}
\end{subequations}
\end{thm}

\begin{thm}\label{thm9}
The asymptotic of $m_{j}$ as $z\rightarrow 0$
\begin{subequations}
\begin{align}
&m_{1}(x,t,z)=\left(
\begin{array}{c}
\mathbf{0}\\
\frac{i}{zq_{0}}[\mathbf{q}_{-}^{\dag}\mathbf{q}_{+}]
\end{array}
\right)+o(1), \quad m_{2}(x,t,z)=\left(
\begin{array}{c}
\mathbf{0}\\
\frac{i}{zq_{0}}[\mathbf{q}_{-}^{\dag}\mathbf{q}_{+}]
\end{array}
\right)+o(1),\\
&m_{3}(x,t,z)=\left(
\begin{array}{c}
-\frac{i\mathbf{q}_{-}}{z}\\
0
\end{array}
\right)+o(1), \quad m_{4}(x,t,z)=\left(
\begin{array}{c}
-\frac{i\mathbf{q}_{+}}{z}\\
0
\end{array}
\right)+o(1)
\end{align}
\end{subequations}
\end{thm}

\noindent\textbf{Proof.}\ \
For convenient, the block diagonal and block off-diagonal terms of $3\times3$ matrix $S$
\begin{equation}
S_{bd}=\left(
\begin{array}{ccc}
s_{11}&s_{12}&0\\
s_{21}&s_{22}&0\\
0&0&s_{33}
\end{array}
\right),\quad
S_{bo}=\left(
\begin{array}{ccc}
0&0&s_{13}\\
0&0&s_{23}\\
s_{31}&s_{32}&0
\end{array}
\right),\quad
[S_{bd}]_{o}=\left(
\begin{array}{ccc}
0&s_{12}&0\\
s_{21}&0&0\\
0&0&0
\end{array}
\right).\nonumber
\end{equation}
And we note that for any matrix $S_{1}$ and $S_{2}$ holds the following qualities
\begin{subequations}
\begin{align}
&[UV]_{bd}=U_{bd}V_{bd}+U_{bo}V_{bo},\quad\quad\quad[UV]_{bo}=U_{bd}V_{bo}+U_{bo}V_{bd},\nonumber\\
&[U_{bd}V_{bd}]_{d}=U_{d}V_{d}+[U_{bd}]_{o}[V_{bd}]_{o},\quad[U_{bd}V_{bd}]_{o}=U_{d}[V_{bd}]_{o}+[U_{bd}]_{o}V_{d}.\nonumber
\end{align}
\end{subequations}
Substituting (\ref{expan}) into the system (\ref{mulax}) and compare the coefficients of $z^{j}$, we deriving the following results
\begin{subequations}
\begin{align}
&E_{0}^{(12)}=E_{0}^{(13)}=E_{0}^{(21)}=E_{0}^{(31)}=E_{0}^{(23)}=E_{0}^{(32)}=0,\nonumber\\
&E_{0}^{(11)}=E_{0}^{(22)}=E_{0}^{(33)}=1,\nonumber\\
&E_{1}^{(13)}=\frac{i}{q_{0}}[\bar{q}q_{\pm}+q\bar{q}_{\pm}-q_{0}^{2}], \quad E_{1}^{(23)}=\frac{i}{q_{0}}[\bar{q}q_{\pm}-q\bar{q}_{\pm}],\nonumber\\
&E_{1}^{(31)}=\frac{i}{q_{0}}[\bar{q}q_{\pm}+q\bar{q}_{\pm}-q_{0}^{2}], \quad E_{1}^{(32)}=\frac{i}{q_{0}}[q\bar{q}_{\pm}-\bar{q}q_{\pm}].\nonumber\\
&F_{0}^{(12)}=F_{0}^{(13)}=F_{0}^{(31)}=F_{0}^{(32)}=F_{1}^{(12)}=F_{1}^{(32)}=0,\nonumber\\
&F_{0}^{(11)}=F_{0}^{(22)}=F_{0}^{(33)}=1,\nonumber\\
&F_{0}^{(21)}=\frac{\bar{q}q_{\pm}-\bar{q}_{\pm}q}{q_{0}^{2}}, \quad F_{0}^{'(23)}=\frac{q^{2}\bar{q}_{\pm}^{2}-q_{\pm}^{2}\bar{q}}{q_{0}^{3}},\nonumber\\
&F_{1}^{(13)}=F_{1}^{(31)}=-\frac{i}{q_{0}^{3}}[\bar{q}q_{\pm}+\bar{q}_{\pm}q-q_{0}^{2}].\nonumber
\end{align}
\end{subequations}
$\Box$

\begin{thm}\label{thm10}
The asymptotic of $a_{jj}$ as $z\rightarrow\infty$
\begin{subequations}
\begin{align}
&a_{11}=\frac{\mathbf{q}_{+}^{\dag}\mathbf{q}_{-}}{q_{0}^{2}}+o(\frac{1}{z}),\quad a_{22}=\frac{\mathbf{q}_{+}^{\dag}\mathbf{q}_{-}}{q_{0}^{2}}+o(\frac{1}{z}),\quad a_{33}=1+o(\frac{1}{z}),\\
&b_{11}=\frac{\mathbf{q}_{-}^{\dag}\mathbf{q}_{+}}{q_{0}^{2}}+o(\frac{1}{z}),\quad b_{22}=\frac{\mathbf{q}_{-}^{\dag}\mathbf{q}_{+}}{q_{0}^{2}}+o(\frac{1}{z}),\quad b_{33}=1+o(\frac{1}{z}),
\end{align}
\end{subequations}
\end{thm}

\begin{thm}\label{thm11}
The asymptotic of $a_{jj}$ as $z\rightarrow 0$
\begin{subequations}
\begin{align}
&a_{11}=1+o(z),\quad a_{22}=\frac{\mathbf{q}_{-}^{\dag}\mathbf{q}_{+}}{q_{0}^{2}}+o(z),\quad a_{33}=\frac{\mathbf{q}_{-}^{\dag}\mathbf{q}_{+}}{q_{0}^{2}}+o(z),\\
&b_{11}=1+o(z),\quad b_{22}=\frac{\mathbf{q}_{-}^{\dag}\mathbf{q}_{+}}{q_{0}^{2}}+o(z),\quad b_{33}=\frac{\mathbf{q}_{-}^{\dag}\mathbf{q}_{+}}{q_{0}^{2}}+o(z),
\end{align}
\end{subequations}
\end{thm}

\section{ Distribution of Discrete Spectral}
For the $3\times 3$ matrix spectral problem, the characters of discrete spectral is more complicate than $2\times 2$ matrix spectral problem. For convenient, we firstly introduce the following $3\times3$ matrix:
\begin{subequations}
\begin{align}
&\Xi_{1}(z)=(\chi_{1}(z), \Phi_{-,2}(z), \Phi_{+,3}(z)), \quad\quad z\in D_{1}\nonumber\\
&\Xi_{2}(z)=(\chi_{2}(z), \Phi_{+,2}(z), \Phi_{-,3}(z)), \quad\quad z\in D_{2}\nonumber\\
&\Xi_{3}(z)=(\Phi_{-,1}(z), \Phi_{+,2}(z), \chi_{3}(z)), \quad\quad z\in D_{3}\nonumber\\
&\Xi_{4}(z)=(\Phi_{+,1}(z), \Phi_{-,2}(z), \chi_{4}(z)), \quad\quad z\in D_{4}\nonumber
\end{align}
\end{subequations}
Then the determinant of $\Xi_{j}(z)$ can be derived via some explicit calculation:
\begin{subequations}
\begin{align}
&\mathrm{Wr}\Xi_{1}(z)=a_{22}(z)b_{33}(z)\gamma e^{i\theta_{2}}, \quad \mathrm{Wr}\Xi_{2}(z)=a_{33}(z)b_{22}(z)\gamma e^{i\theta_{2}},\nonumber\\
&\mathrm{Wr}\Xi_{3}(z)=a_{11}(z)b_{22}(z)\gamma e^{i\theta_{2}}, \quad \mathrm{Wr}\Xi_{4}(z)=a_{22}(z)b_{11}(z)\gamma e^{i\theta_{2}}.\nonumber
\end{align}
\end{subequations}
The columns of $\Xi_{1}(z)$ is linearly dependent at the zeros of $a_{22}(z)$ and $b_{33}(z)$. In the similar way, the other equations can be obtain the similarly properties. For the existence of symmetries of scattering coefficient, we can known that the zeros is dependent of each other.
\begin{lemma}\label{lemma12}
Let $\mathrm{Im}z_{0}>0$, then
\begin{equation}
a_{22}(z_{0})=0\Longleftrightarrow b_{22}(\bar{z}_{0})=0\Longleftrightarrow b_{22}(-\frac{q_{0}^{2}}{\bar{z}_{0}})=0\Longleftrightarrow a_{22}(-\frac{q_{0}^{2}}{z_{0}})=0.\nonumber
\end{equation}
\end{lemma}
\begin{lemma}\label{lemma13}
Let $\mathrm{Im}z_{0}>0$ and $|z_{0}|\geq q_{0}$ then
\begin{equation}
b_{33}(z_{0})=0\Longleftrightarrow a_{33}(\bar{z}_{0})=0\Longleftrightarrow a_{11}(-\frac{q_{0}^{2}}{\bar{z}_{0}})=0\Longleftrightarrow b_{11}(-\frac{q_{0}^{2}}{z_{0}})=0.\nonumber
\end{equation}
\end{lemma}

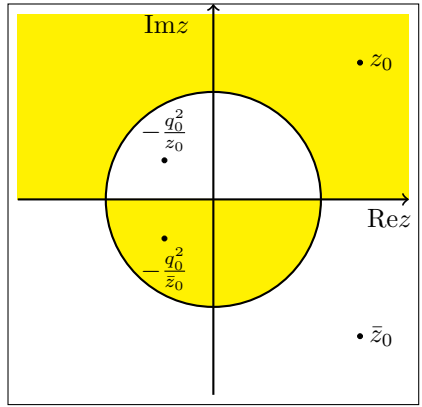
\begin{figure}[H]
\begin{center}
\begin{tikzpicture}[scale=0.65]
\draw[ ](-4.2,-4.2)rectangle(4.2,4);
\path[fill=yellow] (-4,0)rectangle (4,3.8);
\filldraw[white](0,0)--(2.2,0) arc (0:180:2.2);
\filldraw[yellow](0,0)--(-2.2,0) arc (180:360:2.2);
\draw[thick,->] (-4,0)--(4,0);
\draw[ ](3.6,0)node[below] {$\mathrm{Re}z$};
\draw[thick,->](0,-4)--(0,4);
\draw[ ](-0.3,3.6)node[left]{$\mathrm{Im}z$};
\draw [thick] (0,0) circle [radius=2.2];
\draw [fill] (3,2.8) circle [radius=0.05] node[right] {$z_{0}$};
\draw [fill] (3,-2.8) circle [radius=0.05] node[right] {$\bar{z}_{0}$};
\draw [fill] (-1,0.8) circle [radius=0.05] node[above] {$-\frac{q_{0}^{2}}{z_{0}}$};
\draw [fill] (-1,-0.8) circle [radius=0.05] node[below] {$-\frac{q_{0}^{2}}{\bar{z}_{0}}$};
\end{tikzpicture}
\end{center}
\caption{The distribution of discrete spectral.}
\end{figure}

From Lemma\ref{lemma12} and Lemma\ref{lemma13} we can come to the conclusion that the discrete eigenvalues appear in the four kind points: $z_{0}$, $\bar{z}_{0}$, $-\frac{q_{0}^{2}}{z_{0}}$, $-\frac{q_{0}^{2}}{\bar{z}_{0}}$. The distribution of discrete spectral can be shown in Figure 2. It is similar with the $2\times 2$ matrix spectral problem. So, it is enough to just study the zeros of $a_{22}(z)$ and $b_{33}(z)$. The zeros are divided into the following three types:
\begin{itemize}
\item[$\blacktriangle$] {the first kind eigenvalue: $a_{22}(z_{0})=0$ and $b_{33}(z_{0})\neq 0$.}
\item[$\blacktriangle$] {the second kind eigenvalue: $a_{22}(z_{0})\neq0$ and $b_{33}(z_{0})=0$.}
\item[$\blacktriangle$] {the third kind eigenvalue: $a_{22}(z_{0})=0$ and $b_{33}(z_{0})=0$.}
\end{itemize}
Then some results can be obtained:
\begin{lemma}\label{lemma14}
For $\mathrm{Im}z_{0}$ and $|z_{0}|>q_{0}$, the following results are equivalent\\
$\mathrm{(1)}$ $\chi_{1}(z_{0})=0$;\nonumber\\
$\mathrm{(2)}$ $\chi_{4}(-\frac{q_{0}^{2}}{z_{0}})=0$;\nonumber\\
$\mathrm{(3)}$ $\exists b_{0}$, so that $\Phi_{+,2}(\bar{z}_{0})=b_{0}\Phi_{-,3}(\bar{z}_{0})$;\nonumber\\
$\mathrm{(4)}$ $\exists \tilde{b}_{0}$, so that $\Phi_{+,2}(-\frac{q_{0}^{2}}{\bar{z}_{0}})=\tilde{b}_{0}\Phi_{-,1}(-\frac{q_{0}^{2}}{\bar{z}_{0}})$;\nonumber
\end{lemma}
\noindent\textbf{Proof.}\\
$(1)\Longleftrightarrow(2)$:
From (\ref{symchi14}), it can be proofed.\\
$(1)\Longleftrightarrow(3)$:
From (\ref{chi}) and (\ref{barphi}), it can be derived.\\
$(1)\Longleftrightarrow(4)$:
From (\ref{phi123}), it can be obtained. \\

  In the similar way, we have
\begin{lemma}\label{lemma15}
For $\mathrm{Im}z_{0}$ and $|z_{0}|>q_{0}$, the following results are equivalent\\
$\mathrm{(1)}$ $\chi_{2}(\bar{z}_{0})=0$;\nonumber\\
$\mathrm{(2)}$ $\chi_{3}(-\frac{q_{0}^{2}}{\bar{z}_{0}})=0$;\nonumber\\
$\mathrm{(3)}$ $\exists \hat{b}_{0}$, so that $\Phi_{-,2}(z_{0})=\hat{b}_{0}\Phi_{+,3}(z_{0})$;\nonumber\\
$\mathrm{(4)}$ $\exists \check{b}_{0}$, so that $\Phi_{-,2}(-\frac{q_{0}^{2}}{z_{0}})=\check{b}_{0}\Phi_{+,1}(-\frac{q_{0}^{2}}{z_{0}})$;\nonumber
\end{lemma}

\begin{lemma}\label{lemma16}
For $\mathrm{Im}z_{0}$ and $|z_{0}|>q_{0}$, $a_{22}(z_{0})b_{33}(z_{0})=0$, \\
$\mathrm{(1)}$ If $z_{0}$ is the first kind eigenvalue, then
\begin{subequations}
\begin{align}
&\Phi_{-,2}(z_{0})=h_{0}\chi_{1}(z_{0}), \quad\quad\quad\chi_{2}(\bar{z}_{0})=\hat{h}_{0}\Phi_{+,2}(\bar{z}_{0}),\nonumber\\
&\chi_{3}(-\frac{q_{0}^{2}}{\bar{z}_{0}})=\check{h}_{0}\Phi_{+,2}(-\frac{q_{0}^{2}}{\bar{z}_{0}}), \quad \Phi_{-,2}(-\frac{q_{0}^{2}}{z_{0}})=\tilde{h}_{0}\chi_{4}(-\frac{q_{0}^{2}}{z_{0}}).\nonumber
\end{align}
\end{subequations}
$\mathrm{(2)}$ If $z_{0}$ is the second kind eigenvalue, then
\begin{subequations}
\begin{align}
&\chi_{1}(z_{0})=f_{0}\Phi_{+,3}(z_{0}), \quad\quad\quad\Phi_{-,3}(\bar{z}_{0})=\hat{f}_{0}\chi_{2}(\bar{z}_{0}),\nonumber\\
&\Phi_{-,1}(-\frac{q_{0}^{2}}{\bar{z}_{0}})=\check{f}_{0}\chi_{3}(-\frac{q_{0}^{2}}{\bar{z}_{0}}), \quad \chi_{4}(-\frac{q_{0}^{2}}{z_{0}})=\tilde{f}_{0}\Phi_{+,1}(-\frac{q_{0}^{2}}{z_{0}}).\nonumber
\end{align}
\end{subequations}
$\mathrm{(3)}$ If $z_{0}$ is the third kind eigenvalue, then $\chi_{1}(z_{0})=\chi_{2}(\bar{z}_{0})=0$ and
\begin{subequations}
\begin{align}
&\Phi_{-,2}(z_{0})=g_{0}\Phi_{+,3}(z_{0}), \quad \Phi_{-,3}(\bar{z}_{0})=\hat{g}_{0}\Phi_{+,2}(\bar{z}_{0}),\nonumber\\
& \Phi_{-,1}(-\frac{q_{0}^{2}}{\bar{z}_{0}})=\check{g}_{0}\Phi_{+,2}(-\frac{q_{0}^{2}}{\bar{z}_{0}})\quad \Phi_{-,2}(-\frac{q_{0}^{2}}{z_{0}})=\tilde{g}_{0}\Phi_{+,1}(-\frac{q_{0}^{2}}{z_{0}}).\nonumber
\end{align}
\end{subequations}
\end{lemma}

\noindent\textbf{Proof.}\ \
If $a_{22}(z_{0})=0$ and $b_{33}(z_{0})\neq0$, we can derived $\Phi_{-,2}(z_{0})$ and $\chi_{1}(z_{0})$ are linear dependent. So there exist a constant value $h_{0}$, such that $\Phi_{-,2}(z_{0})=h_{0}\chi_{1}(z_{0})$. The rest equations of Lemma\ref{lemma16} can be proofed in the similar way. $\Box$

For convenient when deriving the residue conditions, Lemma\ref{lemma16} can be rewrite in terms of the modified eigenfunctions.\\
(1)Let $\{z_{n}|n=1,2,\cdots N_{1}\}$ are the first kind eigenvalues
\begin{subequations}
\begin{align}
&\mu_{-,2}(z_{n})=h_{n}e^{i(\theta_{1}-\theta_{2})(z_{n})}m_{1}(z_{n}), \quad\quad\quad m_{2}(\bar{z}_{n})=\hat{h}_{n}e^{-i(\theta_{1}-\theta_{2})(\bar{z}_{n})}\mu_{+,2}(\bar{z}_{n}),\nonumber\\
&m_{3}(-\frac{q_{0}^{2}}{\bar{z}_{n}})=\check{h}_{n}e^{i(\theta_{1}+\theta_{2})(-\frac{q_{0}^{2}}{\bar{z}_{n}})}\mu_{+,2}(-\frac{q_{0}^{2}}{\bar{z}_{n}}), \quad \mu_{-,2}(-\frac{q_{0}^{2}}{z_{n}})=\tilde{h}_{n}e^{-i(\theta_{1}+\theta_{2})(-\frac{q_{0}^{2}}{z_{n}})}m_{4}(-\frac{q_{0}^{2}}{z_{n}}).\nonumber
\end{align}
\end{subequations}
(2)If $\{\zeta_{n}|n=1,2,\cdots N_{2}\}$ is the second kind eigenvalues, then
\begin{subequations}
\begin{align}
&m_{1}(\zeta_{n})=f_{n}e^{-2i\theta_{1}(\zeta_{n})}\mu_{+,3}(\zeta_{n}), \quad\quad\mu_{-,3}(\bar{\zeta}_{n})=\hat{f}_{n}e^{2i\theta_{1}(\bar{\zeta}_{n})}m_{2}(\bar{\zeta}_{n}),\nonumber\\
&\mu_{-,1}(-\frac{q_{0}^{2}}{\bar{\zeta}_{n}})=\check{f}_{n}e^{-2i\theta_{1}(-\frac{q_{0}^{2}}{\bar{\zeta}_{n}})}m_{3}(-\frac{q_{0}^{2}}{\bar{\zeta}_{n}}), \quad m_{4}(-\frac{q_{0}^{2}}{\zeta_{n}})=\tilde{f}_{n}e^{2i\theta_{1}(-\frac{q_{0}^{2}}{\zeta_{n}})}\mu_{+,1}(-\frac{q_{0}^{2}}{\zeta_{n}}).\nonumber
\end{align}
\end{subequations}
(3)If $\{\omega_{n}|n=1,2,\cdots N_{3}\}$ is the third kind eigenvalues
\begin{subequations}
\begin{align}
&\mu_{-,2}(\omega_{n})=g_{n}e^{-i(\theta_{1}+\theta_{2})(\omega_{n})}\mu_{+,3}(\omega_{n}),\quad \mu_{-,3}(\bar{\omega}_{n})=\hat{g}_{n}e^{i(\theta_{1}+\theta_{2})(\bar{\omega}_{n})}\mu_{+,2}(\bar{\omega}_{n}),\nonumber\\ &\mu_{-,1}(-\frac{q_{0}^{2}}{\bar{\omega}_{n}})=\check{g}_{n}e^{-i(\theta_{1}-\theta_{2})(-\frac{q_{0}^{2}}{\bar{\omega}_{n}})}\mu_{+,2}(-\frac{q_{0}^{2}}{\bar{\omega}_{n}}),
\quad \mu_{-,2}(-\frac{q_{0}^{2}}{\omega_{n}})=\tilde{g}_{n}e^{i(\theta_{1}-\theta_{2})(-\frac{q_{0}^{2}}{\omega_{n}})}\mu_{+,1}(-\frac{q_{0}^{2}}{\omega_{n}}).\nonumber
\end{align}
\end{subequations}

\section{Riemann-Hilbert Problem}
In this section, we transform solving the asymptotic matrix spectral problem into solving an appropriate RH problem. So we need construct a jump condition that is similar with $2\times 2$. The difference is that the jump matrixes are constructed four times. For convenient, we make some expressions that $D_{+}=D_{1}\cup D_{3}$ and $D_{-}=D_{2}\cup D_{4}$.
\begin{thm}\label{thm12}
The meromorphic function $M(x,t,z)=M_{j}(x,t,z)$, $z\in D_{j}(j=1,2,3,4)$
\[
M^{+}(x,t,z)=\begin{cases}
M_{1}(x,t,z)=\left(\frac{m_{1}}{b_{33}},\frac{\mu_{-,2}}{a_{22}},\mu_{+,3}\right), &z\in D_{1}\\ M_{3}(x,t,z)=\left(\frac{\mu_{-,1}}{a_{11}},\mu_{+,2},\frac{m_{3}}{b_{22}}\right), &z\in D_{3}
\end{cases}
\]
\[
M^{-}(x,t,z)=\begin{cases}
M_
{2}(x,t,z)=\left(\frac{m_{2}}{b_{22}},\mu_{+,2},\frac{\mu_{-,3}}{a_{33}}\right),\quad z\in D_{2}\\
M_{4}(x,t,z)=\left(\mu_{+,1},\frac{\mu_{-,2}}{a_{22}},\frac{m_{4}}{b_{11}}\right),\quad z\in D_{4}
\end{cases}
\]
and $M_{j}(x,t,z)$ hold the jump conditions
\begin{equation}
M^{+}(x,t,z)=M^{-}(x,t,z)[I-e^{i\Theta(x,t,z)}L(z)e^{-i\Theta(x,t,z)}], \quad z\in\Sigma\nonumber
\end{equation}
where
$\Sigma=\Sigma_{1}\cup\Sigma_{2}\cup\Sigma_{3}\cup\Sigma_{4}$, here $\Sigma_{j}=\bar{D}_{j}\cap\bar{D}_{j+1}(j=1,2,3,4)$.
The matrix $L(z)$ are given as
\begin{subequations}
\begin{align}
&L(z)=\left(
\begin{array}{ccc}
-\rho_{2}(-\frac{q_{0}^{2}}{z})\bar{\rho}_{2}(-\frac{q_{0}^{2}}{\bar{z}})&\rho_{3}(z)\rho_{2}(-\frac{q_{0}^{2}}{z})-\frac{z}{iq_{0}}\rho_{3}(-\frac{q_{0}^{2}}{z})&\rho_{2}(-\frac{q_{0}^{2}}{z})\\
l_{1}&l_{2}&l_{3}\\
\bar{\rho}_{2}(-\frac{q_{0}^{2}}{\bar{z}})&-\rho_{3}(z)&0
\end{array}
\right),\quad z\in\Sigma_{1},\\
&L(z)=\left(
\begin{array}{ccc}
\rho_{2}(z)\rho_{2}(-\frac{q_{0}^{2}}{z})&0&\rho_{2}(-\frac{q_{0}^{2}}{z})\\
l_{4}&0&l_{5}\\
\rho_{2}(z)&0&0
\end{array}
\right),\quad z\in\Sigma_{2},\\
&L(z)=\left(
\begin{array}{ccc}
l_{6}&l_{7}&l_{8}\\
-\rho_{1}(z)&0&\bar{\rho}_{3}(\bar{z})\gamma(z)\\
-\rho_{2}(z)+\rho_{1}(z)\rho_{3}(z)&\rho_{3}(z)&-\gamma{z}\bar{\rho}_{3}(\bar{z})\rho_{3}(z)
\end{array}
\right),\quad z\in\Sigma_{3},\\
&L(z)=\left(
\begin{array}{ccc}
\bar{\rho}_{2}(-\frac{q_{0}^{2}}{\bar{z}})\bar{\rho}_{2}(\bar{z})&0&-\bar{\rho}_{2}(\bar{z})\\
0&0&0\\
\bar{\rho}_{2}(-\frac{q_{0}^{2}}{\bar{z}})&0&0
\end{array}
\right),\quad z\in\Sigma_{4},
\end{align}
\end{subequations}
where
\begin{subequations}
\begin{align}
&l_{1}=-\left(\gamma(z)\frac{z}{-iq_{0}}(\bar{\rho}_{3}(-\frac{q_{0}^{2}}{\bar{z}})+\rho_{2}(-\frac{q_{0}^{2}}{z})\bar{\rho}_{2}(-\frac{q_{0}^{2}}{\bar{z}})\bar{\rho}_{3}(-\frac{q_{0}^{2}}{\bar{z}}))+\bar{\rho}_{2}(-\frac{q_{0}^{2}}{\bar{z}})\rho_{1}(-\frac{q_{0}^{2}}{z})\frac{iq_{0}}{z}\right),\nonumber\\
&l_{2}=\gamma(z)\frac{z}{-iq_{0}}\left(\rho_{3}(z)\rho_{2}(-\frac{q_{0}^{2}}{z})\bar{\rho}_{3}(-\frac{q_{0}^{2}}{\bar{z}})-\frac{z}{iq_{0}}\rho_{3}(-\frac{q_{0}^{2}}{z})\bar{\rho}_{3}(-\frac{q_{0}^{2}}{\bar{z}})\right)+\frac{iq_{0}}{z}\rho_{3}(z)\rho_{1}(-\frac{q_{0}^{2}}{z}),\nonumber\\
&l_{3}=\gamma(z)\frac{z}{-iq_{0}}\rho_{2}(-\frac{q_{0}^{2}}{z})\bar{\rho}_{3}(-\frac{q_{0}^{2}}{\bar{z}})+\frac{iq_{0}}{z}\rho_{1}(-\frac{q_{0}^{2}}{z}),\nonumber\\
&l_{4}=\gamma(z)\frac{z}{-iq_{0}}(\rho_{2}(z)\rho_{2}(-\frac{q_{0}^{2}}{z})\bar{\rho}_{3}(-\frac{q_{0}^{2}}{\bar{z}})-\bar{\rho}_{3}(-\frac{q_{0}^{2}}{\bar{z}}))+\frac{iq_{0}}{z}\rho_{2}(z)\rho_{1}(-\frac{q_{0}^{2}}{z})-\rho_{1}(z),\nonumber\\
&l_{5}=\gamma(z)\bar{\rho}_{3}(-\frac{q_{0}^{2}}{\bar{z}})+\gamma(z)\frac{z}{-iq_{0}}\rho_{2}(-\frac{q_{0}^{2}}{z})\bar{\rho}_{3}(-\frac{q_{0}^{2}}{\bar{z}})+\frac{iq_{0}}{z}\rho_{1}(-\frac{q_{0}^{2}}{z}),\nonumber\\
&l_{6}=\frac{z}{iq_{0}}\rho_{1}(z)\rho_{3}(-\frac{q_{0}^{2}}{z})-\rho_{2}(z)\bar{\rho}_{2}(\bar{z})+\rho_{1}(z)\rho_{3}(z)\bar{\rho}_{2}(\bar{z}),\nonumber\\
&l_{7}=\frac{z}{iq_{0}}\rho_{3}(-\frac{q_{0}^{2}}{z})+\rho_{3}(z)\bar{\rho}_{2}(\bar{z}),\nonumber\\
&l_{8}=-\gamma(z)\frac{z}{iq_{0}}\bar{\rho}_{3}(\bar{z})\rho_{3}(-\frac{q_{0}^{2}}{z})-\gamma(z)\rho_{3}(z)\bar{\rho}_{3}(\bar{z})\bar{\rho}_{2}(\bar{z})-\bar{\rho}_{2}(\bar{z}).\nonumber
\end{align}
\end{subequations}
\end{thm}

\noindent\textbf{Proof.}\ \
From (\ref{phia}) and Theorem \ref{thm6}, we can obtain
\begin{subequations}
\begin{align}
&\frac{\Phi_{-,2}}{a_{22}}=\left(\frac{a_{12}}{a_{22}}-\frac{a_{32}}{a_{22}}\frac{a_{13}}{a_{33}}\right)\frac{\chi_{2}}{b_{22}}
+\left(1+\frac{a_{12}}{a_{22}}\frac{b_{21}}{b_{22}}-\frac{a_{32}}{a_{22}}\frac{a_{13}}{a_{33}}\frac{b_{21}}{b_{22}}-\frac{a_{32}}{a_{22}}\frac{a_{23}}{a_{33}}\right)\Phi_{+,2}
+\frac{a_{32}}{a_{22}}\frac{\Phi_{-,3}}{a_{33}},\nonumber\\
&\Phi_{+,3}=-\frac{a_{13}}{a_{33}}\frac{\chi_{2}}{b_{22}}+\left(-\frac{a_{13}}{a_{33}}\frac{b_{21}}{b_{22}}-\frac{a_{23}}{a_{33}}\right)\Phi_{+,2}+\frac{\Phi_{-,3}}{a_{33}},\nonumber\\
&\frac{\chi_{1}}{b_{33}}=\left(1+\frac{b_{31}}{b_{33}}\frac{a_{13}}{a_{33}}\right)\frac{\chi_{2}}{b_{22}}
+\left(\frac{b_{21}}{b_{22}}+\frac{b_{31}}{b_{33}}\frac{a_{13}}{a_{33}}\frac{b_{21}}{b_{22}}+\frac{b_{31}}{b_{33}}\frac{a_{23}}{a_{33}}\right)\Phi_{+,2}-\frac{b_{31}}{b_{33}}\frac{\Phi_{-,3}}{a_{33}}.\nonumber
\end{align}
\end{subequations}
Then we can obtain the matrix $L(z)$ as $z\in\Sigma_{1}$. The others can be derived in the similar way. $\Box$

From the asymptotic behavior of $\mu_{\pm}$, $m_{j}$ and scattering coefficient in Section 5, the asymptotic behavior of $M(x,t,z)$ can be obtained:
\begin{equation}
M(x,t,z)=\left(
\begin{array}{ccc}
-\frac{\mathbf{q}_{+}}{q_{0}}&\frac{\mathbf{q}_{+}^{\perp}}{q_{0}}&0\\
0&0&1
\end{array}
\right)+o(\frac{1}{z}),\quad z\rightarrow\infty,
\end{equation}
\begin{equation}
M(x,t,z)=\left(
\begin{array}{ccc}
0&0&-\frac{i\mathbf{q}_{+}}{z}\\
\frac{iq_{0}}{z}&0&0
\end{array}
\right)+o(1),\quad z\rightarrow 0,
\end{equation}
For convenient, we define the notation $M^{\pm}=(m^{\pm}_{1},m^{\pm}_{2},m^{\pm}_{3})$, and $M^{\pm}_{-1,\nu}(x,t)$ represent the residue of $M^{\pm}$ at $z=\nu$.
\begin{thm}\label{thm13}
The meromorphic functions $M^{\pm}$ satisfy the residue conditions
\begin{subequations}
\begin{align}
&M^{+}_{-1,z_{n}}=\left(0,H_{n}m_{1}^{+}(z_{n})e^{i(\theta_{1}-\theta_{2})(z_{n})},0\right),\\
&M^{-}_{-1,\bar{z}_{n}}=\left(\hat{H}_{n}m_{2}^{-}(\bar{z}_{n})e^{-i(\theta_{1}-\theta_{2})(\bar{z}_{n})},0,0\right)\\
&M^{+}_{-1,-\frac{q_{0}^{2}}{\bar{z}_{n}}}=\left(0,0,\check{H}_{n}m_{2}^{-}(\bar{z}_{n})e^{i(\theta_{1}+\theta_{2})(-\frac{q_{0}^{2}}{\bar{z}_{n}})}\right),\\ &M^{-}_{-1,-\frac{q_{0}^{2}}{z_{n}}}=\left(0,\frac{z_{n}}{iq_{0}}\tilde{H}_{n}m_{1}^{+}(z_{n})e^{-i(\theta_{1}+\theta_{2})(-\frac{q_{0}^{2}}{z_{n}})},0\right)\\
&M^{+}_{-1,\zeta_{n}}=\left(F_{n}m_{3}^{+}(\zeta_{n})e^{-2i\theta_{1}(\zeta_{n})},0,0\right),\\ &M^{-}_{-1,\bar{\zeta}_{n}}=\left(0,0,\hat{F}_{n}m_{1}^{-}(\bar{\zeta}_{n})e^{2i\theta_{1}(\bar{\zeta}_{n})}\right)\\
&M^{+}_{-1,-\frac{q_{0}^{2}}{\bar{\zeta}_{n}}}=\left(\frac{\bar{\zeta}_{n}}{iq_{0}}\check{F}_{n}m_{1}^{-}(\bar{\zeta}_{n })e^{-2i\theta_{1}(-\frac{q_{0}^{2}}{\bar{\zeta}_{n}})},0,0\right)\\ &M^{-}_{-1,-\frac{q_{0}^{2}}{\zeta_{n}}}=\left(0,0,\frac{\zeta_{n}}{iq_{0}}\tilde{F}_{n}m_{3}^{+}(\zeta_{n})e^{2i\theta_{1}(-\frac{q_{0}^{2}}{\zeta_{n}})}\right)\\
&M^{+}_{-1,\omega_{n}}=\left(0,G_{n}m_{3}^{+}(\omega_{n})e^{-i(\theta_{1}+\theta_{2})(\omega_{n})},0\right),\\ &M^{-}_{-1,\bar{\omega}_{n}}=\left(0,0,\hat{G}_{n}m_{2}^{-}(\bar{\omega}_{n})e^{i(\theta_{1}+\theta_{2})(\bar{\omega}_{n})}\right)\\
&M^{+}_{-1,-\frac{q_{0}^{2}}{\bar{\omega}_{n}}}=\left(\check{G}_{n}m_{2}^{-}({\bar{\omega}_{n}})e^{-i(\theta_{1}-\theta_{2})(-\frac{q_{0}^{2}}{\bar{\omega}_{n}})},0,0\right)\\ &M^{-}_{-1,-\frac{q_{0}^{2}}{\omega_{n}}}=\left(0,\frac{\omega_{n}}{iq_{0}}\tilde{G}_{n}m_{3}^{+}(\omega_{n})e^{i(\theta_{1}-\theta_{2})(-\frac{q_{0}^{2}}{\omega_{n}})},0\right)
\end{align}\label{res}
\end{subequations}
where
\begin{subequations}
\begin{align}
&H_{n}=\frac{h_{n}b_{33}(z_{n})}{a'_{22}(z_{n})},\quad \hat{H}_{n}=\frac{\hat{h}_{n}}{b'_{22}(\bar{z}_{n})},\quad\check{H}_{n}=\frac{\check{h}_{n}}{b'_{22}(-\frac{q_{0}^{2}}{\bar{z}_{n}})},\quad
\tilde{H}_{n}=\frac{\tilde{h}_{n}b_{33}(z_{n})}{a'_{22}(-\frac{q_{0}^{2}}{z_{n}})}\\
&F_{n}=\frac{f_{n}}{b'_{33}(\zeta_{n})},\quad \hat{F}_{n}=\frac{\hat{f}_{n}b_{22}(\bar{\zeta}_{n})}{a'_{33}(\bar{\zeta}_{n})},\quad\check{F}_{n}=\frac{\check{f}_{n}b_{22}(\bar{\zeta}_{n})}{a'_{11}(-\frac{q_{0}^{2}}{\bar{\zeta}_{n}})},\quad
\tilde{F}_{n}=\frac{\tilde{f}_{n}}{b'_{11}(-\frac{q_{0}^{2}}{\zeta_{n}})}\\
&G_{n}=\frac{g_{n}}{a'_{22}(\omega_{n})},\quad \hat{G}_{n}=\frac{\hat{g}_{n}}{a'_{33}(\bar{\omega}_{n})},\quad \check{G}_{n}=\frac{\check{g}_{n}}{a'_{11}(-\frac{q_{0}^{2}}{\bar{\omega}_{n}})},\quad
\tilde{G}_{n}=\frac{\tilde{g}_{n}}{a'_{22}(-\frac{q_{0}^{2}}{\omega_{n}})}
\end{align}
\end{subequations}
\end{thm}
\begin{lemma}\label{lemma17}
The constants in Theorem\ref{lemma16} obey the following relations
\begin{subequations}
\begin{align}
&\tilde{h}_{n}=\frac{iq_{0}}{z_{n}}h_{n},\quad \hat{h}_{n}=\frac{iq_{0}}{\bar{z}_{n}}\check{h}_{n}=-\gamma(\bar{z}_{n})a_{33}(\bar{z}_{n})\bar{h}_{n}\\
&\tilde{f}_{n}=f_{n},\quad \hat{f}_{n}=\check{f}_{n}=-\frac{\bar{f}_{n}}{\bar{a}_{22}(\zeta_{n})}\\
&\tilde{g}_{n}=\frac{iq_{0}}{\omega_{n}}g_{n},\quad \hat{g}_{n}=\frac{iq_{0}}{\bar{\omega}_{n}}\check{g}_{n}=-\frac{a'_{33}(\bar{\omega}_{n})}{b'_{22}(\bar{\omega}_{n})}\gamma(\bar{\omega}_{n})\bar{g}_{n}.
\end{align}
\end{subequations}
\end{lemma}
\begin{lemma}\label{lemma18}
The constants in Theorem\ref{thm13} obey the following relations
\begin{subequations}
\begin{align}
&\hat{H}_{n}=-\gamma(\bar{z}_{n})\bar{H}_{n}, \quad \check{H}_{n}=\frac{iq_{0}}{\bar{z}_{n}}\gamma(\bar{z}_{n})\bar{H}_{n},\quad \tilde{H}_{n}=-\frac{z_{n}}{iq_{0}}H_{n},\\
&\hat{F}_{n}=-\bar{F}_{n},\quad \check{F}_{n}=-\frac{q_{0}^{2}}{\bar{\zeta}_{n}^{2}}\bar{F}_{n},\quad \tilde{F}_{n}=\frac{q_{0}^{2}}{\zeta_{n}^{2}}F_{n},\\
&\hat{G}_{n}=-\gamma(\bar{\omega}_{n})\bar{G}_{n},\quad \check{G}_{n}=-\frac{\bar{\omega}_{n}^{3}}{iq_{0}^{3}}\gamma(\bar{\omega}_{n})\bar{G}_{n},\quad \tilde{G}_{n}=-\frac{\omega_{n}}{iq_{0}}G_{n}
\end{align}
\end{subequations}
\end{lemma}

\noindent\textbf{Proof of Lemma\ref{lemma18}}\\
From equation (\ref{phit}) and the first symmetry of Jost eigenfunctions
\begin{equation}
(\Phi^{\dag}_{\pm}(x,t,\bar{z}))^{-1}C(z)=\Phi_{\pm}(x,t,z),
\end{equation}
we obtain
\begin{subequations}
\begin{align}
&\Phi_{-,3}(\bar{z})=\frac{\gamma(\bar{z})}{\det{\bar{\Phi}_{-}(z)}}[\bar{\Phi}_{-,1}(z)\times\bar{\Phi}_{-,2}(z)],\\
&\Phi_{+,2}(\bar{z})=\frac{1}{\det{\bar{\Phi}_{+}(z)}}[\bar{\Phi}_{+,3}(z)\times\bar{\Phi}_{+,1}(z)].
\end{align}
\end{subequations}
Substituting them into the first two equations of (3) in Lemma \ref{lemma16} and combine the relation (\ref{phia}), we deriving the result
\begin{equation}
-g_{0}\bar{\gamma}(\bar{z}_{0})[a_{11}(z_{0})(\Phi_{+,3}(z_{0})\times\Phi_{+,1}(z_{0}))+a_{21}(z_{0})(\Phi_{+,3}(z_{0})\times\Phi_{+,2}(z_{0}))]
=\bar{\hat{g}}_{0}\Phi_{+,3}(z_{0})\times\Phi_{+,1}(z_{0}).\label{phig0}
\end{equation}
One can dot product the vector $\Phi_{+,2}(z_{0})$ on the both sides of equation (\ref{phig0}).
 Then $\hat{g}_{0}=-\bar{g}_{0}\gamma(\bar{z}_{0})\bar{a}_{11}(z_{0})$.

\subsection{Trace Formula}
 By $B=A^{-1}$ and $\det{A}=\det{B}=1$, one can obtain
\begin{equation}
a_{22}b_{22}=1/[1+\frac{a_{12}b_{21}+a_{32}b_{23}}{a_{22}b_{22}}].\label{ab22}
\end{equation}
Furthermore, taking logarithmic of both sides of (\ref{ab22}) and combined with the reflection coefficients (\ref{rho}), we can get
\begin{equation}
\log{a_{22}(z)}-\log{\frac{1}{b_{22}(z)}}=J_{0}(z).\label{log}
\end{equation}
where
\begin{equation}
J_{0}(z)=\log[1+\gamma(z)\rho_{3}(-\frac{q_{0}^{2}}{z})\bar{\rho}_{3}(-\frac{q_{0}^{2}}{\bar{z}})+\gamma(z)\rho_{3}(z)\bar{\rho}_{3}(\bar{z})].\nonumber
\end{equation}
Since $a_{22}$ and $b_{22}$ are analytic in the upper-half plane and lower-half plane, equation (\ref{log}) is a jump condition for a scalar, additive Riemann-Hilbert problem. To circumvent the pole singularities coming from the zeros of $a_{22}$ and $b_{22}$, we define two analytic function
\begin{subequations}
\begin{align}
&\beta^{+}(z)=a_{22}(z)e^{i\Delta\theta}\prod_{n=1}^{N_{1}}\frac{z-\bar{z}_{n}}{z-z_{n}}\frac{z-(-\frac{q_{0}^{2}}{\bar{z}_{n}})}{z-(-\frac{q_{0}^{2}}{z_{n}})}
\prod_{n=1}^{N_{3}}\frac{z-\bar{\omega}_{n}}{z-\omega_{n}}\frac{z-(-\frac{q_{0}^{2}}{\bar{\omega}_{n}})}{z-(-\frac{q_{0}^{2}}{\omega_{n}})},\quad z\in\mathbb{C}^{+}\\
&\beta^{-}(z)=\frac{1}{b_{22}}(z)e^{i\Delta\theta}\prod_{n=1}^{N_{1}}\frac{z-\bar{z}_{n}}{z-z_{n}}\frac{z-(-\frac{q_{0}^{2}}{\bar{z}_{n}})}{z-(-\frac{q_{0}^{2}}{z_{n}})}
\prod_{n=1}^{N_{3}}\frac{z-\bar{\omega}_{n}}{z-\omega_{n}}\frac{z-(-\frac{q_{0}^{2}}{\bar{\omega}_{n}})}{z-(-\frac{q_{0}^{2}}{\omega_{n}})},\quad z\in\mathbb{C}^{-}
\end{align}
\end{subequations}
where $\Delta\theta=\theta_{+}-\theta_{-}$,
and $\beta^{\pm}(z)$ no zeros in $\mathbb{C}^{\pm}$ and approaches 1 as $z\rightarrow\infty$ in the analytic region in $z$-plane. Using the Plemej's formula and the projectors $P^{\pm}$ we obtain $\log\beta(z)=P(\log[\beta^{+}\beta^{-}])$ for $z\in\mathbb{C}\setminus\Sigma$,
where $P$ is the Cauchy projectors
\begin{equation}
P_{\pm}(f)(z)=\frac{1}{2\pi i}\int_{\Sigma}\frac{f(\xi)}{\xi-(z\pm i0)}\mathrm{\xi}.
\end{equation}
Combine (\ref{log}) and taking the exponentials,
\begin{equation}
a_{22}(z)=\exp\left(-i\Delta\theta-\frac{1}{2\pi i}\int_{\mathbb{R}}\frac{J_{0}(\xi)}{\xi-z}\mathrm{d}{\xi}\right)
\prod_{n=1}^{N_{1}}\frac{z-z_{n}}{z-\bar{z}_{n}}\frac{z-(-\frac{q_{0}^{2}}{z_{n}})}{z-(-\frac{q_{0}^{2}}{\bar{z}_{n}})}
\prod_{n=1}^{N_{3}}\frac{z-\omega_{n}}{z-\bar{\omega}_{n}}\frac{z-(-\frac{q_{0}^{2}}{\omega_{n}})}{z-(-\frac{q_{0}^{2}}{\bar{\omega}_{n}})}\label{a11}
\end{equation}
In the similar way as above yields
\begin{subequations}
\begin{align}
&\log b_{33}(z)-\log\frac{1}{a_{33}(z)}=-\log\left[1+\frac{1}{\gamma(z)}\rho_{1}(-\frac{q_{0}^{2}}{z})\bar{\rho}_{1}(-\frac{q_{0}^{2}}{\bar{z}})
+\rho_{2}(-\frac{q_{0}^{2}}{z})\bar{\rho}_{2}(-\frac{q_{0}^{2}}{\bar{z}})\right],\\
&\log a_{11}(z)-\log\frac{1}{b_{11}(z)}=-\log\left[1+\gamma(z)\rho_{1}(z)\bar{\rho}_{1}(\bar{z})
+\rho_{2}(z)\bar{\rho}_{2}(\bar{z})\right],
\end{align}\label{logab1}
\end{subequations}
However, $b_{33}(z)$, $a_{33}(z)$, $a_{11}(z)$ and $b_{11}(z)$ are only analytic in $D_{1}$, $D_{2}$, $D_{3}$, $D_{4}$. This is the reason that this situation is complicated than above.To formula a appropriate Riemann-Hilbert problem, we have to introduce a sectionally function which is analytic on the whole $z$-plane.
So we obtain
\begin{subequations}
\begin{align}
&\log{b_{33}(z)}-\log{\frac{1}{b_{11}(z)}}=\log{a_{22}(z)}-\log{[1-\bar{\rho}_{2}(\bar{z})\bar{\rho}_{2}(-\frac{q_{0}^{2}}{\bar{z}})]},\\
&\log{a_{11}(z)}-\log{\frac{1}{a_{33}(z)}}=\log{b_{22}(z)}-\log{[1-\rho_{2}(z)\rho_{2}(-\frac{q_{0}^{2}}{z})]},
\end{align}\label{logab2}
\end{subequations}
Next, we define the following sectionally analytic function
\[
\hat{\beta}^{+}(z)=\begin{cases}
\beta_{1}(z), &z\in D_{1}\\
\beta_{3}(z), &z\in D_{3}
\end{cases}\quad
\hat{\beta}^{-}(z)=\begin{cases}
\beta_{2}(z), &z\in D_{2}\\
\beta_{4}(z), &z\in D_{4}
\end{cases}
\]
where
\begin{subequations}
\begin{align}
&\beta_{1}(z)=\frac{b_{33}(z)}{s_{1}(z)}\prod_{n=1}^{N_{1}}\frac{z-z_{n}}{z-\bar{z}_{n}}\frac{z-(-\frac{q_{0}^{2}}{\bar{z}_{n}})}{z-(-\frac{q_{0}^{2}}{z_{n}})}
\prod_{n=1}^{N_{2}}\frac{z-\bar{\zeta}_{n}}{z-\zeta_{n}}\frac{z-(-\frac{q_{0}^{2}}{\zeta_{n}})}{z-(-\frac{q_{0}^{2}}{\bar{\zeta}_{n}})},\\
&\beta_{2}(z)=\frac{\bar{s}_{1}(\bar{z})}{a_{33}(z)}\prod_{n=1}^{N_{1}}\frac{z-z_{n}}{z-\bar{z}_{n}}\frac{z-(-\frac{q_{0}^{2}}{\bar{z}_{n}})}{z-(-\frac{q_{0}^{2}}{z_{n}})}
\prod_{n=1}^{N_{2}}\frac{z-\bar{\zeta}_{n}}{z-\zeta_{n}}\frac{z-(-\frac{q_{0}^{2}}{\zeta_{n}})}{z-(-\frac{q_{0}^{2}}{\bar{\zeta}_{n}})},\\
&\beta_{3}(z)=\frac{a_{11}(z)}{\bar{s}_{2}(\bar{z})}\prod_{n=1}^{N_{1}}\frac{z-z_{n}}{z-\bar{z}_{n}}\frac{z-(-\frac{q_{0}^{2}}{\bar{z}_{n}})}{z-(-\frac{q_{0}^{2}}{z_{n}})}
\prod_{n=1}^{N_{2}}\frac{z-\bar{\zeta}_{n}}{z-\zeta_{n}}\frac{z-(-\frac{q_{0}^{2}}{\zeta_{n}})}{z-(-\frac{q_{0}^{2}}{\bar{\zeta}_{n}})},\\
&\beta_{4}(z)=\frac{s_{2}(z)}{b_{11}(z)}\prod_{n=1}^{N_{1}}\frac{z-z_{n}}{z-\bar{z}_{n}}\frac{z-(-\frac{q_{0}^{2}}{\bar{z}_{n}})}{z-(-\frac{q_{0}^{2}}{z_{n}})}
\prod_{n=1}^{N_{2}}\frac{z-\bar{\zeta}_{n}}{z-\zeta_{n}}\frac{z-(-\frac{q_{0}^{2}}{\zeta_{n}})}{z-(-\frac{q_{0}^{2}}{\bar{\zeta}_{n}})},
\end{align}
\end{subequations}
where
\begin{equation}
s_{1}(z)=\prod_{n=1}^{N_{1}}\frac{z-z_{n}}{z-\bar{z}_{n}}\prod_{n=1}^{N_{3}}\frac{z-\omega_{n}}{z-\bar{\omega}_{n}},\quad
s_{2}(z)=\prod_{n=1}^{N_{1}}\frac{z-(-\frac{q_{0}^{2}}{z_{n}})}{z-(-\frac{q_{0}^{2}}{\bar{z}_{n}})}\prod_{n=1}^{N_{3}}\frac{z-(-\frac{q_{0}^{2}}{\omega_{n}})}{z-(-\frac{q_{0}^{2}}{\bar{\omega}_{n}})},
\end{equation}
Then (\ref{logab1}) can be rewritten as
\begin{equation}
\log \beta_{1}(z)-\log\beta_{2}(z)=J_{1}(z),\quad\log \beta_{3}(z)-\log\beta_{4}(z)=J_{3}(z),
\label{jump12}
\end{equation}
where
\begin{subequations}
\begin{align}
&J_{1}(z)=-\log\left[1+\frac{1}{\gamma(z)}\rho_{1}(-\frac{q_{0}^{2}}{z})\bar{\rho}_{1}(-\frac{q_{0}^{2}}{\bar{z}})
+\rho_{2}(-\frac{q_{0}^{2}}{z})\bar{\rho}_{2}(-\frac{q_{0}^{2}}{\bar{z}})\right],\\
&J_{3}(z)=-\log\left[1+\gamma(z)\rho_{1}(z)\bar{\rho}_{1}(\bar{z})
+\rho_{2}(z)\bar{\rho}_{2}(\bar{z})\right],
\end{align}
\end{subequations}
Combine $(\ref{a11})$ and $\beta_{j}(z)$, (\ref{logab2}) can be rewritten as
\begin{equation}
\log{\beta_{3}(z)}-\log{\beta_{2}(z)}=J_{2}(z),\quad\log{\beta_{1}(z)}-\log{\beta_{4}(z)}=J_{4}(z),
\label{jump34}
\end{equation}
where
\begin{subequations}
\begin{align}
&J_{2}(z)=\frac{1}{2\pi i}\log\int_{\mathbb{R}}\frac{\bar{J}_{0}(\xi)}{\xi-z}\mathrm{d}\xi-\log{[1-\bar{\rho}_{2}(\bar{z})\bar{\rho}_{2}(-\frac{q_{0}^{2}}{\bar{z}})]},\\
&J_{4}(z)=-\frac{1}{2\pi i}\log\int_{\mathbb{R}}\frac{J_{0}(\xi)}{\xi-z}\mathrm{d}\xi-\log{[1-\rho_{2}(z)\rho_{2}(-\frac{q_{0}^{2}}{z})]},
\end{align}
\end{subequations}

Now, (\ref{jump12}) and (\ref{jump34}) form the jump conditions for the Riemann-Hilbert problem with the analytic functions $\hat{\beta}^{\pm}(z)$. It can be precisely rewritten as
\begin{equation}
\log{\hat{\beta}^{+}(z)}-\log{\hat{\beta}^{-}(z)}=J_{i}(z), \quad J_{i}(z)=\Sigma_{i}(i=1,2,3,4).
\end{equation}
Applying the projectors $P^{\pm}$ and Plemelj's formula yields
\begin{equation}
\log{\hat{\beta}(z)}=\frac{1}{2\pi i}\int_{\Sigma}\frac{J(\xi)}{\xi-z}\mathrm{d}\xi,\quad z\in\mathbb{C}\backslash\Sigma
\end{equation}
Taking the exponential of both sides with $z\in D_{1}$yields
\begin{equation}
\frac{b_{33}(z)}{s_{1}(z)}=\exp\left(\frac{1}{2\pi i}\int_{\Sigma}\frac{J(\xi)}{\xi-z}\mathrm{d}\xi\right)\prod_{n=1}^{N_{1}}\frac{z-\bar{z}_{n}}{z-z_{n}}\frac{z-(-\frac{q_{0}^{2}}{z_{n}})}{z-(-\frac{q_{0}^{2}}{\bar{z}_{n}})}
\prod_{n=1}^{N_{2}}\frac{z-\zeta_{n}}{z-\bar{\zeta}_{n}}\frac{z-(-\frac{q_{0}^{2}}{\bar{\zeta}_{n}})}{z-(-\frac{q_{0}^{2}}{\zeta_{n}})},
\end{equation}
Next, we obtain
\begin{equation}
b_{33}(z)=\exp\left(\frac{1}{2\pi i}\int_{\Sigma}\frac{J(\xi)}{\xi-z}\mathrm{d}\xi\right)\prod_{n=1}^{N_{1}}\frac{z-(-\frac{q_{0}^{2}}{z_{n}})}{z-(-\frac{q_{0}^{2}}{\bar{z}_{n}})}
\prod_{n=1}^{N_{2}}\frac{z-\zeta_{n}}{z-\bar{\zeta}_{n}}\frac{z-(-\frac{q_{0}^{2}}{\bar{\zeta}_{n}})}{z-(-\frac{q_{0}^{2}}{\zeta_{n}})}\prod_{n=1}^{N_{3}}\frac{z-\omega_{n}}{z-\bar{\omega}_{n}},
\end{equation}
The asymptotic phase difference $\Delta\theta=\theta_{+}-\theta_{-}$ is given by
\begin{equation}
\Delta\theta=\frac{1}{2\pi}\int_{\Sigma}\frac{J(\xi)}{\xi}\mathrm{\xi}+2\sum_{n=1}^{N_{1}}\mathrm{arg}z_{n}-4\sum_{n=1}^{N_{2}}\mathrm{arg}\zeta_{n}-2\sum_{n=1}^{N_{3}}\mathrm{arg}\omega_{n}
\end{equation}

\section{Solution of Riemann-Hilbert Problem}
\begin{thm}\label{thm14}
Let $\nu_{n}$ is the set of all discrete spectral, the solution of the RH problem in Theorem \ref{thm12} is given by
\begin{equation}
M(x,t,z)=\Gamma_{+}(z)+\sum_{n=1}^{N}(\frac{M^{+}_{-1,\nu_{n}}}{z-\nu_{n}}+\frac{M^{-}_{-1,\bar{\nu}_{n}}}{z-\bar{\nu}_{n}}+\frac{M^{+}_{-1,-\frac{q_{0}^{2}}{\bar{\nu}_{n}}}}{z-(-\frac{q_{0}^{2}}{\bar{\nu}_{n}})}+\frac{M^{-}_{-1,-\frac{q_{0}^{2}}{\nu_{n}}}}{z-(-\frac{q_{0}^{2}}{\nu_{n}})})-\frac{1}{2\pi i}\int_{\Sigma}\frac{M^{-}(\xi)\hat{L}(\xi)}{\xi-z}\mathrm{d}\xi,
\end{equation}
where $N=N_{1}+N_{2}+N_{3}$, $\hat{L}=e^{i\Theta}Le^{-i\Theta}$. The modified eigenfunctions in the residue conditions (\ref{res}) can be given as
\begin{subequations}
\begin{align}
&\begin{split}m^{-}_{2}(z)=&
\begin{pmatrix}
\frac{\mathbf{q}_{+}^{\perp}}{q_{0}}\\0
\end{pmatrix}+\sum_{n=1}^{N_{1}}\left(\frac{H_{n}e^{i(\theta_{1}-\theta_{2})(z_{n})}}{z-z_{n}}+\frac{z_{n}}{iq_{0}}\frac{\tilde{H}_{n}e^{-i(\theta_{1}+\theta_{2})(-\frac{q_{0}^{2}}{z_{n}})}}{z-(-\frac{q_{0}^{2}}{z_{n}})}\right)m_{1}^{+}(z_{n})\\
&+\sum_{n=1}^{N_{3}}\left(\frac{G_{n}e^{-i(\theta_{1}+\theta_{2})(\omega_{n})}}{z-\omega_{n}}+\frac{\omega_{n}}{iq_{0}}\frac{\tilde{G}_{n}e^{i(\theta_{1}-\theta_{2})(-\frac{q_{0}^{2}}{\omega_{n}})}}{z-(-\frac{q_{0}^{2}}{\omega_{n}})}\right)m_{3}^{+}(\omega_{n})-\frac{1}{2\pi i}\int_{\Sigma}\frac{(M^{-}\hat{L}(\xi))_{2}}{\xi-z}\mathrm{d}\xi
\end{split},\\
&\begin{split}m^{+}_{1}(z)=&
\begin{pmatrix}
-\frac{\mathbf{q}_{+}}{q_{0}}\\\frac{iq_{0}}{z}
\end{pmatrix}+\sum_{n=1}^{N_{1}}\frac{\hat{H}_{n}e^{-i(\theta_{1}-\theta_{2})(\bar{z}_{n})}}{z-\bar{z}_{n}}m_{2}^{-}(\bar{z}_{n})
+\sum_{n=1}^{N_{2}}\frac{F_{n}e^{-2i\theta_{1}(\zeta_{n})}}{z-{\zeta_{n}}}m_{3}^{+}(\zeta_{n})\\
&+\sum_{n=1}^{N_{2}}\frac{\bar{\zeta}_{n}}{iq_{0}}\frac{\check{F}_{n}e^{-2i\theta_{1}(-\frac{q_{0}^{2}}{\bar{\zeta}_{n}})}}{z-(-\frac{q_{0}^{2}}{\bar{\zeta}_{n}})}m_{1}^{-}(\bar{\zeta}_{n})
+\sum_{n=1}^{N_{3}}\frac{\check{G}_{n}e^{-i(\theta_{1}-\theta_{2})(-\frac{q_{0}^{2}}{\bar{\omega}_{n}})}}{z-(-\frac{q_{0}^{2}}{\bar{\omega}_{n}})}m_{2}^{-}(\bar{\omega}_{n})-\frac{1}{2\pi i}\int_{\Sigma}\frac{(M^{-}\hat{L}(\xi))_{1}}{\xi-z}\mathrm{d}\xi
\end{split}\\
&\begin{split}m^{-}_{1}(z)=&
\begin{pmatrix}
-\frac{\mathbf{q}_{+}}{q_{0}}\\\frac{iq_{0}}{z}
\end{pmatrix}+\sum_{n=1}^{N_{1}}\frac{\hat{H}_{n}e^{-i(\theta_{1}-\theta_{2})(\bar{z}_{n})}}{z-\bar{z}_{n}}m_{2}^{-}(\bar{z}_{n})
+\sum_{n=1}^{N_{2}}\frac{F_{n}e^{-2i\theta_{1}(\zeta_{n})}}{z-{\zeta_{n}}}m_{3}^{+}(\zeta_{n})\\
&+\sum_{n=1}^{N_{2}}\frac{\bar{\zeta}_{n}}{iq_{0}}\frac{\check{F}_{n}e^{-2i\theta_{1}(-\frac{q_{0}^{2}}{\bar{\zeta}_{n}})}}{z-(-\frac{q_{0}^{2}}{\bar{\zeta}_{n}})}m_{1}^{-}(\bar{\zeta}_{n})
+\sum_{n=1}^{N_{3}}\frac{\check{G}_{n}e^{-i(\theta_{1}-\theta_{2})(-\frac{q_{0}^{2}}{\bar{\omega}_{n}})}}{z-(-\frac{q_{0}^{2}}{\bar{\omega}_{n}})}m_{2}^{-}(\bar{\omega}_{n})-\frac{1}{2\pi i}\int_{\Sigma}\frac{(M^{-}\hat{L}(\xi))_{1}}{\xi-z}\mathrm{d}\xi
\end{split}\\
&\begin{split}m^{+}_{3}(z)=&
\begin{pmatrix}
-\frac{i\mathbf{q}_{+}}{z}\\1
\end{pmatrix}+\sum_{n=1}^{N_{1}}\frac{\check{H}_{n}e^{i(\theta_{1}+\theta_{2})(-\frac{q_{0}^{2}}{\bar{z}_{n}})}}{z-(-\frac{q_{0}^{2}}{\bar{z}_{n}})}m_{2}^{-}(\bar{z}_{n})
+\sum_{n=1}^{N_{2}}\frac{\hat{F}_{n}e^{2i\theta_{1}(\bar{\zeta}_{n})}}{z-{\bar{\zeta}_{n}}}m_{1}^{-}(\bar{\zeta}_{n})\\
&+\sum_{n=1}^{N_{2}}\frac{\zeta_{n}}{iq_{0}}\frac{\tilde{F}_{n}e^{2i\theta_{1}(-\frac{q_{0}^{2}}{\zeta_{n}})}}{z-(-\frac{q_{0}^{2}}{\zeta_{n}})}m_{3}^{+}(\zeta_{n})
+\sum_{n=1}^{N_{3}}\frac{\hat{G}_{n}e^{i(\theta_{1}+\theta_{2})(\bar{\omega}_{n})}}{z-{\bar{\omega}_{n}}}m_{2}^{-}(\bar{\omega}_{n})-\frac{1}{2\pi i}\int_{\Sigma}\frac{(M^{-}\hat{L}(\xi))_{3}}{\xi-z}\mathrm{d}\xi
\end{split}\end{align}
\end{subequations}
\end{thm}

\begin{thm}\label{thm15}
The solution of Riemann-Hilbert problem in Theorem \ref{thm14} is reconstructed as
\begin{equation}
\begin{split}
q(x,t)&=q_{+}+i\sum_{n=1}^{N_{1}}\check{H}_{n}e^{i(\theta_{1}+\theta_{2})(-\frac{q_{0}^{2}}{\bar{z}_{n}})}m_{12}^{-}(\bar{z}_{n})
+i\sum_{n=1}^{N_{2}}\frac{\zeta_{n}}{iq_{0}}\tilde{F}_{n}e^{2i\theta_{1}(-\frac{q_{0}^{2}}{\zeta_{n}})}m_{13}^{+}(\zeta_{n})\\
&+i\sum_{n=1}^{N_{2}}\hat{F}_{n}e^{2i\theta_{1}(\bar{\zeta}_{n})}m_{11}^{-}(\bar{\zeta}_{n})+i\sum_{n=1}^{N_{3}}\hat{G}_{n}e^{i(\theta_{1}+\theta_{2})(\bar{\omega}_{n})}m_{12}^{-}(\bar{\omega}_{n})
\end{split}\label{q}
\end{equation}
\end{thm}

\begin{subequations}
\begin{align}
&\begin{split}m^{-}_{12}(\bar{z}_{i'})=&\frac{q_{+}}{q_{0}}
+\sum_{n=1}^{N_{1}}\left(\frac{H_{n}e^{i(\theta_{1}-\theta_{2})(z_{n})}}{\bar{z}_{i'}-z_{n}}+\frac{z_{n}}{iq_{0}}\frac{\tilde{H}_{n}e^{-i(\theta_{1}+\theta_{2})(-\frac{q_{0}^{2}}{z_{n}})}}{\bar{z}_{i'}-(-\frac{q_{0}^{2}}{z_{n}})}\right)m_{11}^{+}(z_{n})\\
&+\sum_{n=1}^{N_{3}}\left(\frac{G_{n}e^{-i(\theta_{1}+\theta_{2})(\omega_{n})}}{\bar{z}_{i'}-\omega_{n}}+\frac{\omega_{n}}{iq_{0}}\frac{\tilde{G}_{n}e^{i(\theta_{1}-\theta_{2})(-\frac{q_{0}^{2}}{\omega_{n}})}}{\bar{z}_{i'}-(-\frac{q_{0}^{2}}{\omega_{n}})}\right)m_{13}^{+}(\omega_{n})
\end{split},\label{a}\\
&\begin{split}m^{-}_{12}(\bar{\omega}_{l'})=&\frac{q_{+}}{q_{0}}
+\sum_{n=1}^{N_{1}}\left(\frac{H_{n}e^{i(\theta_{1}-\theta_{2})(z_{n})}}{\bar{\omega}_{l'}-z_{n}}+\frac{z_{n}}{iq_{0}}\frac{\tilde{H}_{n}e^{-i(\theta_{1}+\theta_{2})(-\frac{q_{0}^{2}}{z_{n}})}}{\bar{\omega}_{l'}-(-\frac{q_{0}^{2}}{z_{n}})}\right)m_{11}^{+}(z_{n})\\
&+\sum_{n=1}^{N_{3}}\left(\frac{G_{n}e^{-i(\theta_{1}+\theta_{2})(\omega_{n})}}{\bar{\omega}_{l'}-\omega_{n}}+\frac{\omega_{n}}{iq_{0}}\frac{\tilde{G}_{n}e^{i(\theta_{1}-\theta_{2})(-\frac{q_{0}^{2}}{\omega_{n}})}}{\bar{\omega}_{l'}-(-\frac{q_{0}^{2}}{\omega_{n}})}\right)m_{13}^{+}(\omega_{n})
\end{split},\label{b}
\end{align}

\begin{align}
&\begin{split}m^{+}_{11}(z_{i'})=&-\frac{q_{+}}{q_{0}}
+\sum_{n=1}^{N_{1}}\frac{\hat{H}_{n}e^{-i(\theta_{1}-\theta_{2})(\bar{z}_{n})}}{z_{i'}-\bar{z}_{n}}m_{12}^{-}(\bar{z}_{n})
+\sum_{n=1}^{N_{2}}\frac{F_{n}e^{-2i\theta_{1}(\zeta_{n})}}{z_{i'}-{\zeta_{n}}}m_{13}^{+}(\zeta_{n})\\
&+\sum_{n=1}^{N_{2}}\frac{\bar{\zeta}_{n}}{iq_{0}}\frac{\check{F}_{n}e^{-2i\theta_{1}(-\frac{q_{0}^{2}}{\bar{\zeta}_{n}})}}{z_{i'}-(-\frac{q_{0}^{2}}{\bar{\zeta}_{n}})}m_{11}^{-}(\bar{\zeta}_{n})
+\sum_{n=1}^{N_{3}}\frac{\check{G}_{n}e^{-i(\theta_{1}-\theta_{2})(-\frac{q_{0}^{2}}{\bar{\omega}_{n}})}}{z_{i'}-(-\frac{q_{0}^{2}}{\bar{\omega}_{n}})}m_{12}^{-}(\bar{\omega}_{n})
\end{split}\label{c}\\
&\begin{split}m^{-}_{11}(\bar{\zeta}_{j'})=&-\frac{q_{+}}{q_{0}}
+\sum_{n=1}^{N_{1}}\frac{\hat{H}_{n}e^{-i(\theta_{1}-\theta_{2})(\bar{z}_{n})}}{\bar{\zeta}_{j'}-\bar{z}_{n}}m_{12}^{-}(\bar{z}_{n})
+\sum_{n=1}^{N_{2}}\frac{F_{n}e^{-2i\theta_{1}(\zeta_{n})}}{\bar{\zeta}_{j'}-{\zeta_{n}}}m_{13}^{+}(\zeta_{n})\\
&+\sum_{n=1}^{N_{2}}\frac{\bar{\zeta}_{n}}{iq_{0}}\frac{\check{F}_{n}e^{-2i\theta_{1}(-\frac{q_{0}^{2}}{\bar{\zeta}_{n}})}}{\bar{\zeta}_{j'}-(-\frac{q_{0}^{2}}{\bar{\zeta}_{n}})}m_{11}^{-}(\bar{\zeta}_{n})
+\sum_{n=1}^{N_{3}}\frac{\check{G}_{n}e^{-i(\theta_{1}-\theta_{2})(-\frac{q_{0}^{2}}{\bar{\omega}_{n}})}}{\bar{\zeta}_{j'}-(-\frac{q_{0}^{2}}{\bar{\omega}_{n}})}m_{12}^{-}(\bar{\omega}_{n})
\end{split}\label{d}
\end{align}

\begin{align}
&\begin{split}m^{+}_{13}(\zeta_{j'})=&-\frac{iq_{+}}{\zeta_{j'}}
+\sum_{n=1}^{N_{1}}\frac{\check{H}_{n}e^{i(\theta_{1}+\theta_{2})(-\frac{q_{0}^{2}}{\bar{z}_{n}})}}{\zeta_{j'}-(-\frac{q_{0}^{2}}{\bar{z}_{n}})}m_{12}^{-}(\bar{z}_{n})
+\sum_{n=1}^{N_{2}}\frac{\hat{F}_{n}e^{2i\theta_{1}(\bar{\zeta}_{n})}}{\zeta_{j'}-{\bar{\zeta}_{n}}}m_{11}^{-}(\bar{\zeta}_{n})\\
&+\sum_{n=1}^{N_{2}}\frac{\zeta_{n}}{iq_{0}}\frac{\tilde{F}_{n}e^{2i\theta_{1}(-\frac{q_{0}^{2}}{\zeta_{n}})}}{\zeta_{j'}-(-\frac{q_{0}^{2}}{\zeta_{n}})}m_{13}^{+}(\zeta_{n})
+\sum_{n=1}^{N_{3}}\frac{\hat{G}_{n}e^{i(\theta_{1}+\theta_{2})(\bar{\omega}_{n})}}{\zeta_{j'}-{\bar{\omega}_{n}}}m_{12}^{-}(\bar{\omega}_{n})
\end{split}\label{e}\\
&\begin{split}m^{+}_{13}(\omega_{l'})=&-\frac{iq_{+}}{\omega_{l'}}
+\sum_{n=1}^{N_{1}}\frac{\check{H}_{n}e^{i(\theta_{1}+\theta_{2})(-\frac{q_{0}^{2}}{\bar{z}_{n}})}}{\omega_{l'}-(-\frac{q_{0}^{2}}{\bar{z}_{n}})}m_{12}^{-}(\bar{z}_{n})
+\sum_{n=1}^{N_{2}}\frac{\hat{F}_{n}e^{2i\theta_{1}(\bar{\zeta}_{n})}}{\omega_{l'}-{\bar{\zeta}_{n}}}m_{11}^{-}(\bar{\zeta}_{n})\\
&+\sum_{n=1}^{N_{2}}\frac{\zeta_{n}}{iq_{0}}\frac{\tilde{F}_{n}e^{2i\theta_{1}(-\frac{q_{0}^{2}}{\zeta_{n}})}}{\omega_{l'}-(-\frac{q_{0}^{2}}{\zeta_{n}})}m_{13}^{+}(\zeta_{n})
+\sum_{n=1}^{N_{3}}\frac{\hat{G}_{n}e^{i(\theta_{1}+\theta_{2})(\bar{\omega}_{n})}}{\omega_{l'}-{\bar{\omega}_{n}}}m_{12}^{-}(\bar{\omega}_{n})
\end{split}\label{f}
\end{align}
\end{subequations}
where
$i'=1,2,\cdots,N_{1}$, $j'=1,2,\cdots,N_{2}$ and $l'=1,2,\cdots,N_{3}$.

For convenient, we define the following notations:
\begin{subequations}
\begin{align}
&\Delta_{n}^{(1)}(x,t,z)=\frac{H_{n}e^{i(\theta_{1}-\theta_{2})(z_{n})}}{z-z_{n}}+\frac{z_{n}}{iq_{0}}\frac{\tilde{H}_{n}e^{-i(\theta_{1}+\theta_{2})(-\frac{q_{0}^{2}}{z_{n}})}}{z-(-\frac{q_{0}^{2}}{z_{n}})}\\
&\Delta_{n}^{(2)}(x,t,z)=\frac{G_{n}e^{-i(\theta_{1}+\theta_{2})(\omega_{n})}}{z-\omega_{n}}+\frac{\omega_{n}}{iq_{0}}\frac{\tilde{G}_{n}e^{i(\theta_{1}-\theta_{2})(-\frac{q_{0}^{2}}{\omega_{n}})}}{z-(-\frac{q_{0}^{2}}{\omega_{n}})}\\
&\Delta_{n}^{(3)}(x,t,z)=\frac{\hat{H}_{n}e^{-i(\theta_{1}-\theta_{2})(\bar{z}_{n})}}{z-\bar{z}_{n}},\quad\Delta_{n}^{(4)}(x,t,z)=\frac{F_{n}e^{-2i\theta_{1}(\zeta_{n})}}{z-{\zeta_{n}}}\\
&\Delta_{n}^{(5)}(x,t,z)=\frac{\bar{\zeta}_{n}}{iq_{0}}\frac{\check{F}_{n}e^{-2i\theta_{1}(-\frac{q_{0}^{2}}{\bar{\zeta}_{n}})}}{z-(-\frac{q_{0}^{2}}{\bar{\zeta}_{n}})},\quad\Delta_{n}^{(6)}(x,t,z)=\frac{\check{G}_{n}e^{-i(\theta_{1}-\theta_{2})(-\frac{q_{0}^{2}}{\bar{\omega}_{n}})}}{z-(-\frac{q_{0}^{2}}{\bar{\omega}_{n}})}\\
&\Delta_{n}^{(7)}(x,t,z)=\frac{\check{H}_{n}e^{i(\theta_{1}+\theta_{2})(-\frac{q_{0}^{2}}{\bar{z}_{n}})}}{z-(-\frac{q_{0}^{2}}{\bar{z}_{n}})},\quad\Delta_{n}^{(8)}(x,t,z)=\frac{\hat{F}_{n}e^{2i\theta_{1}(\bar{\zeta}_{n})}}{z-{\bar{\zeta}_{n}}}\\
&\Delta_{n}^{(9)}(x,t,z)=\frac{\zeta_{n}}{iq_{0}}\frac{\tilde{F}_{n}e^{2i\theta_{1}(-\frac{q_{0}^{2}}{\zeta_{n}})}}{z-(-\frac{q_{0}^{2}}{\zeta_{n}})},\quad\Delta_{n}^{(10)}(x,t,z)=\frac{\hat{G}_{n}e^{i(\theta_{1}+\theta_{2})(\bar{\omega}_{n})}}{z-{\bar{\omega}_{n}}}
\end{align}\label{g}
\end{subequations}

Substituting (\ref{c}) and (\ref{f}) into (\ref{a}) and (\ref{b}), we obtain
\begin{equation}
\begin{split}
m_{12}^{-}(z)&=\frac{q_{+}}{q_{0}}-\sum_{n=1}^{N_{1}}\frac{q_{+}}{q_{0}}\Delta_{n}^{(1)}(z)-\sum_{n=1}^{N_{3}}\frac{iq_{+}}{\omega_{n}}\Delta_{n}^{(2)}(z)
+\sum_{n=1}^{N_{1}}\sum_{n'=1}^{N_{2}}\Delta_{n}^{(1)}(z)\Delta_{n'}^{(4)}(z_{n})m_{13}^{+}(\zeta_{n'})\\
&+\sum_{n=1}^{N_{1}}\sum_{n'=1}^{N_{2}}\Delta_{n}^{(1)}(z)\Delta_{n'}^{(5)}(z_{n})m_{11}^{-}(\bar{\zeta}_{n'})
+\sum_{n=1}^{N_{1}}\sum_{n'=1}^{N_{1}}\Delta_{n}^{(1)}(z)\Delta_{n'}^{(3)}(z_{n})m_{12}^{-}(\bar{z}_{n'})\\
&+\sum_{n=1}^{N_{1}}\sum_{n'=1}^{N_{3}}\Delta_{n}^{(1)}(z)\Delta_{n'}^{(6)}(z_{n})m_{12}^{-}(\bar{\omega}_{n'})
+\sum_{n=1}^{N_{3}}\sum_{n'=1}^{N_{2}}\Delta_{n}^{(2)}(z)\Delta_{n'}^{(9)}(\omega_{n})m_{13}^{+}(\zeta_{n'})\\
&+\sum_{n=1}^{N_{3}}\sum_{n'=1}^{N_{2}}\Delta_{n}^{(2)}(z)\Delta_{n'}^{(8)}(\omega_{n})m_{11}^{-}(\bar{\zeta}_{n'})
+\sum_{n=1}^{N_{3}}\sum_{n'=1}^{N_{1}}\Delta_{n}^{(2)}(z)\Delta_{n'}^{(7)}(\omega_{n})m_{12}^{-}(\bar{z}_{n'})\\
&+\sum_{n=1}^{N_{3}}\sum_{n'=1}^{N_{3}}\Delta_{n}^{(2)}(z)\Delta_{n'}^{(10)}(\omega_{n})m_{12}^{-}(\bar{\omega}_{n'}).\label{h}
\end{split}
\end{equation}
where $z=\bar{z}_{i'}$ and $z=\bar{\omega}_{l'}$.
Together with (\ref{d}), (\ref{f}) and (\ref{h})comprise a linear equations.To solve this equations by Gramer's rule, we define the vector
$\mathbf{x}=(x_{1},\cdots,x_{N_{1}+2N_{2}+N_{3}})$,
 \begin{equation}
x_{k}=\left\{
\begin{aligned}
&m_{12}^{-}(\bar{z}_{k}),& &k=1,\cdots,N_{1},\\
&m_{13}^{+}(\zeta_{k-N_{1}}),& &k=N_{1}+1,\cdots,N_{1}+N_{2},\\
&m_{11}^{-}(\bar{\zeta}_{k-N_{1}-N_{2}}),& &k=N_{1}+N_{2}+1,\cdots,N_{1}+2N_{2},\\
&m_{12}^{-}(\bar{\omega}_{k-N_{1}-2N_{2}}),& &k=N_{1}+2N_{2}+1,\cdots,N_{1}+2N_{2}+N_{3},
\end{aligned}\right.\nonumber
\end{equation}
So we obtain a systems equations
\begin{equation}
(\mathbf{I}-\mathbf{F})\mathbf{x}_{k}=\mathbf{W}\mathbf{x}_{k}=\mathbf{b}_{k},\label{if}
\end{equation}
where
 \begin{equation}
b_{k}=\left\{
\begin{aligned}
&\begin{split}\frac{q_{+}}{q_{0}}&-\sum_{n=1}^{N_{1}}\frac{q_{+}}{q_{0}}\Delta_{n}^{(1)}(\bar{z}_{k})\\
&-\sum_{n=1}^{N_{3}}\frac{iq_{+}}{\omega_{n}}\Delta_{n}^{(2)}(\bar{z}_{k}),\end{split}& &k=1,\cdots,N_{1},\\
&-\frac{iq_{+}}{\zeta_{k-N_{1}}},& &k=N_{1}+1,\cdots,N_{1}+N_{2},\\
&-\frac{q_{+}}{q_{0}},& &k=N_{1}+N_{2}+1,\cdots,N_{1}+2N_{2},\\
&\begin{split}\frac{q_{+}}{q_{0}}&-\sum_{n=1}^{N_{1}}\frac{q_{+}}{q_{0}}\Delta_{n}^{(1)}(\bar{\omega}_{k-N_{1}-2N_{2}})\\
&-\sum_{n=1}^{N_{3}}\frac{iq_{+}}{\omega_{n}}\Delta_{n}^{(2)}(\bar{\omega}_{k-N_{1}-2N_{2}}),\end{split}& &k=N_{1}+2N_{2}+1,\cdots,N_{1}+2N_{2}+N_{3},
\end{aligned}\right.\nonumber
\end{equation}

\begin{equation}
y_{k}=\left\{
\begin{aligned}
&-i\check{H}_{k}e^{i(\theta_{1}+\theta_{2})(-\frac{q_{0}^{2}}{\bar{z}_{k}})},& &k=1,\cdots,N_{1},\\
&-\frac{\zeta_{k-N_{1}}}{q_{0}}\tilde{F}_{k-N_{1}}e^{2i\theta_{1}(-\frac{q_{0}^{2}}{\zeta_{k-N_{1}}})},& &k=N_{1}+1,\cdots,N_{1}+N_{2},\\
&-i\hat{F}_{k-N_{1}-N_{2}}e^{2i\theta_{1}(\bar{\zeta}_{k-N_{1}-N_{2}})},& &k=N_{1}+N_{2}+1,\cdots,N_{1}+2N_{2},\\
&-i\hat{G}_{k-N_{1}-2N_{2}}e^{i(\theta_{1}+\theta_{2})(\bar{\omega}_{k-N_{1}-2N_{2}})},& &k=N_{1}+2N_{2}+1,\cdots,N_{1}+2N_{2}+N_{3}.
\end{aligned}\right.
\end{equation}
The matrix $\mathbf{F}$ are defined as follows:\\
For $i,j=1,\cdots,N_{1}$
\begin{equation}
F_{ij}=\sum_{n=1}^{N_{1}}\Delta_{n}^{(1)}(\bar{z}_{i})\Delta_{j}^{(3)}(z_{n})+\sum_{n=1}^{N_{3}}\Delta_{n}^{(2)}(\bar{z}_{i})\Delta_{j}^{(7)}(\omega_{n})
\end{equation}
For $i=1,\cdots,N_{1}$ and $j=N_{1}+1,\cdots,N_{1}+N_{2}$
\begin{equation}
F_{ij}=\sum_{n=1}^{N_{1}}\Delta_{n}^{(1)}(\bar{z}_{i})\Delta_{j-N_{1}}^{(4)}(z_{n})+\sum_{n=1}^{N_{3}}\Delta_{n}^{(2)}(\bar{z}_{i})\Delta_{j-N_{1}}^{(9)}(\omega_{n})
\end{equation}
For $i=1,\cdots,N_{1}$ and $j=N_{1}+N_{2}+1,\cdots,N_{1}+2N_{2}$
\begin{equation}
F_{ij}=\sum_{n=1}^{N_{1}}\Delta_{n}^{(1)}(\bar{z}_{i})\Delta_{j-N_{1}-N_{2}}^{(5)}(z_{n})+\sum_{n=1}^{N_{3}}\Delta_{n}^{(2)}(\bar{z}_{i})\Delta_{j-N_{1}-N_{2}}^{(8)}(\omega_{n})
\end{equation}
For $i=1,\cdots,N_{1}$ and $j=N_{1}+2N_{2}+1,\cdots,N_{1}+2N_{2}+N_{3}$
\begin{equation}
F_{ij}=\sum_{n=1}^{N_{1}}\Delta_{n}^{(1)}(\bar{z}_{i})\Delta_{j-N_{1}-2N_{2}}^{(6)}(z_{n})+\sum_{n=1}^{N_{3}}\Delta_{n}^{(2)}(\bar{z}_{i})\Delta_{j-N_{1}-2N_{2}}^{(10)}(\omega_{n})
\end{equation}
For $i=N_{1}+1,\cdots,N_{2}$ and $j=1,\cdots,N_{1}$
\begin{equation}
F_{ij}=\Delta_{j}^{(7)}(\zeta_{i-N_{1}})
\end{equation}
For $i,j=N_{1}+1,\cdots,N_{2}$
\begin{equation}
F_{ij}=\Delta_{j-N_{1}}^{(9)}(\zeta_{i-N_{1}})
\end{equation}
For $i=N_{1}+1,\cdots,N_{2}$ and $j=N_{1}+N_{2}+1,\cdots,N_{1}+2N_{2}$
\begin{equation}
F_{ij}=\Delta_{j-N_{1}-N_{2}}^{(8)}(\zeta_{i-N_{1}})
\end{equation}
For $i=N_{1}+1,\cdots,N_{2}$ and $j=N_{1}+2N_{2}+1,\cdots,N_{1}+2N_{2}+N_{3}$
\begin{equation}
F_{ij}=\Delta_{j-N_{1}-2N_{2}}^{(10)}(\zeta_{i-N_{1}})
\end{equation}
For $i=N_{1}+N_{2}+1,\cdots,N_{1}+2N_{2}$ and $j=1,\cdots,N_{1}$
\begin{equation}
F_{ij}=\Delta_{j}^{(3)}(\bar{\zeta}_{i-N_{1}-N_{2}})
\end{equation}
For $i=N_{1}+N_{2}+1,\cdots,N_{1}+2N_{2}$ and $j=N_{1}+1,\cdots,N_{1}+N_{2}$
\begin{equation}
F_{ij}=\Delta_{j-N_{1}}^{(4)}(\bar{\zeta}_{i-N_{1}-N_{2}})
\end{equation}
For $i,j=N_{1}+N_{2}+1,\cdots,N_{1}+2N_{2}$
\begin{equation}
F_{ij}=\Delta_{j-N_{1}-N_{2}}^{(5)}(\bar{\zeta}_{i-N_{1}-N_{2}})
\end{equation}
For $i=N_{1}+N_{2}+1,\cdots,N_{1}+2N_{2}$ and $j=N_{1}+2N_{2}+1,\cdots,N_{1}+2N_{2}+N_{3}$
\begin{equation}
F_{ij}=\Delta_{j-N_{1}-2N_{2}}^{(6)}(\bar{\zeta}_{i-N_{1}-N_{2}})
\end{equation}
For $i=N_{1}+2N_{2}+1,\cdots,N_{1}+2N_{2}+N_{3}$ and $j=1,\cdots,N_{1}$
\begin{equation}
F_{ij}=\sum_{n=1}^{N_{1}}\Delta_{n}^{(1)}(\bar{\omega}_{i-N_{1}-2N_{2}})\Delta_{j}^{(3)}(z_{n})+\sum_{n=1}^{N_{3}}\Delta_{n}^{(2)}(\bar{\omega}_{i-N_{1}-2N_{2}})\Delta_{j}^{(7)}(\omega_{n})
\end{equation}
For $i=N_{1}+2N_{2}+1,\cdots,N_{1}+2N_{2}+N_{3}$ and $j=N_{1}+1,\cdots,N_{1}+N_{2}$
\begin{equation}
F_{ij}=\sum_{n=1}^{N_{1}}\Delta_{n}^{(1)}(\bar{\omega}_{i-N_{1}-2N_{2}})\Delta_{j-N_{1}}^{(4)}(z_{n})+\sum_{n=1}^{N_{3}}\Delta_{n}^{(2)}(\bar{\omega}_{i-N_{1}-2N_{2}})\Delta_{j-N_{1}}^{(9)}(\omega_{n})
\end{equation}
For $i=N_{1}+2N_{2}+1,\cdots,N_{1}+2N_{2}+N_{3}$ and $j=N_{1}+N_{2}+1,\cdots,N_{1}+2N_{2}$
\begin{equation}
F_{ij}=\sum_{n=1}^{N_{1}}\Delta_{n}^{(1)}(\bar{\omega}_{i-N_{1}-2N_{2}})\Delta_{j-N_{1}-N_{2}}^{(5)}(z_{n})+\sum_{n=1}^{N_{3}}\Delta_{n}^{(2)}(\bar{\omega}_{i-N_{1}-2N_{2}})\Delta_{j-N_{1}-N_{2}}^{(8)}(\omega_{n})
\end{equation}
For $i,j=N_{1}+2N_{2}+1,\cdots,N_{1}+2N_{2}+N_{3}$
\begin{equation}
F_{ij}=\sum_{n=1}^{N_{1}}\Delta_{n}^{(1)}(\bar{\omega}_{i-N_{1}-2N_{2}})\Delta_{j-N_{1}-2N_{2}}^{(6)}(z_{n})+\sum_{n=1}^{N_{3}}\Delta_{n}^{(2)}(\bar{\omega}_{i-N_{1}-2N_{2}})\Delta_{j-N_{1}-2N_{2}}^{(10)}(\omega_{n})
\end{equation}
By the Gramer's rule, one can derive that the components of the solutions of system (\ref{if})
\begin{equation}
x_{k}=\frac{\det\mathbf{W}_{k}^{aug}}{\det\mathbf{W}},\quad k=1,2,\cdots,N_{1}+2N_{2}+N_{3},
\end{equation}
where $\mathbf{W}_{k}^{aug}=(\mathbf{W}_{1},\cdots,\mathbf{W}_{k-1},\mathbf{b},\mathbf{W}_{k+1},\cdots,\mathbf{W}_{N_{1}+2N_{2}+N_{3}})$
Substituting $x_{k}$ into the reconstruction formula (\ref{q}) yields the potential function
\begin{equation}
q(x,t)=\frac{\det\mathbf{W}^{aug}}{\det\mathbf{W}}
\end{equation}
\begin{equation}
\mathbf{W}^{aug}=\left(\begin{array}{cc}
q_{+}&\mathbf{y}^{T}\\
\mathbf{b}&\mathbf{W}
\end{array}\right)
\end{equation}
where
\begin{equation}
\mathbf{b}=(b_{1},\cdots,b_{N_{1}+2N_{2}+N_{3}}),\quad \mathbf{y}=(y_{1},\cdots,y_{N_{1}+2N_{2}+N_{3}})^{T}
\end{equation}

\section{ The Single Soliton Solutions }
\subsection{Case I }
In this subsection, we assume the eigenvalues $\nu_{n}$ is the first kind eigenvalues. This implies that $N_{1}=1$ and $N_{2}=N_{3}=0$. And we also assume that
$\mathbf{q}_{+}=(1,1)^{T}\quad H_{1}=e^{\alpha+i\eta},\quad z=\rho e^{i\delta},(0<\delta<\pi)$.
So we derived the soliton solution
\begin{equation}
q(x,t)=\frac{\det\left(
\begin{array}{cc}
q_{+}&y_{1}\\
b_{1}&1-F_{11}
\end{array}
\right)}{1-F_{11}},\nonumber
\end{equation}
where
\begin{subequations}
\begin{align}
&F_{11}=\Delta_{1}^{(1)}(\bar{z}_{1})\Delta_{1}^{(3)}(z_{1}),\quad b_{1}=\frac{q_{+}}{q_{0}}-\frac{q_{+}}{q_{0}}\Delta_{1}^{(1)}(\bar{z}_{1}),\quad y_{1}=-i\check{H}_{1}e^{i(\theta_{1}+\theta_{2})(-\frac{q_{0}^{2}}{\bar{z}_{1}})},\nonumber\\
&\Delta_{1}^{(1)}(\bar{z}_{1})=\frac{H_{1}e^{i(\theta_{1}-\theta_{2})(z_{1})}}{\bar{z}_{1}-z_{1}}+\frac{z_{1}}{iq_{0}}\frac{\tilde{H}_{1}e^{-i(\theta_{1}+\theta_{2})(-\frac{q_{0}^{2}}{z_{1}})}}{\bar{z}_{1}-(-\frac{q_{0}^{2}}{z_{1}})},\nonumber\\
&\Delta_{1}^{(3)}(z_{1})=\frac{\hat{H}_{1}e^{-i(\theta_{1}-\theta_{2})(\bar{z}_{1})}}{z_{1}-\bar{z}_{1}}.\nonumber
\end{align}
\end{subequations}
Figure 3-4 showing the perspective view and propagation view of the one soliton solution as different parameters. Figure 3 showing the situation of $\rho<q_{0}$. Figure 4 showing $\rho>q_{0}$.

\begin{center}
{\includegraphics[scale=0.35]{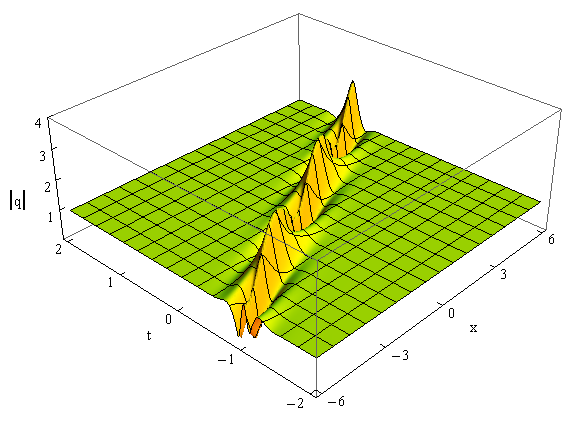}}\hspace{0.5cm}{\includegraphics[scale=0.35]{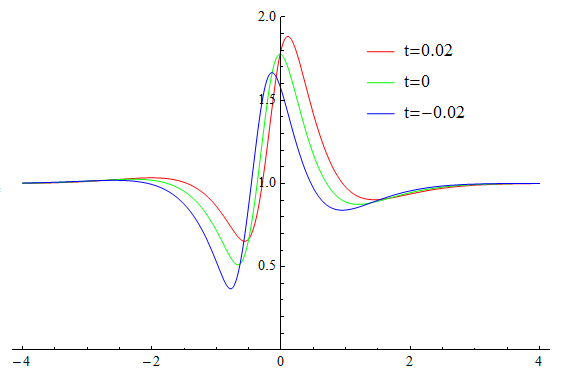}}\\
\vspace{-0.2cm}{\footnotesize\hspace{0.5cm}(a)\hspace{4.5cm}(b)}\\
 \flushleft{\footnotesize {\bf Fig.~3.} $q_{0}=\sqrt{2}$, $\rho=1$, $\delta=\frac{\pi}{3}$, $\alpha=0$, $\eta=-\frac{\pi}{2}$. (a) the perspective view of one soliton. (b) the propagation view of soliton solution in different time.}
\end{center}

\begin{center}
{\includegraphics[scale=0.35]{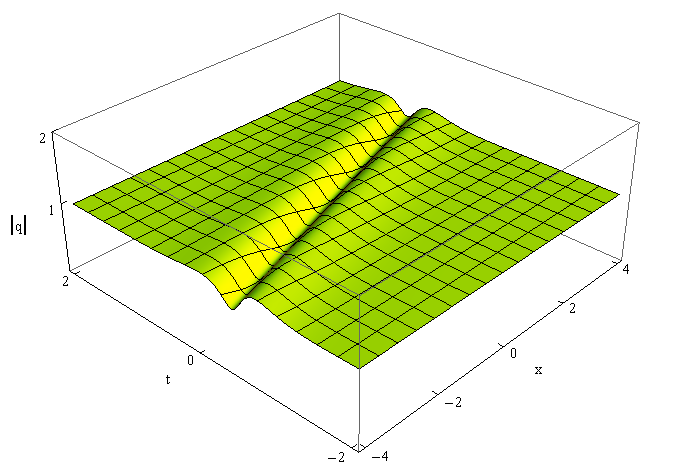}}\hspace{0.5cm}{\includegraphics[scale=0.35]{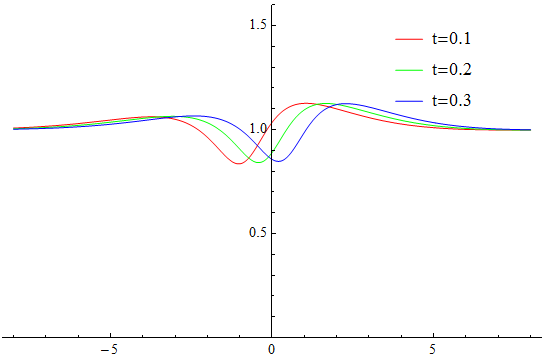}}\\
\vspace{-0.2cm}{\footnotesize\hspace{0.5cm}(a)\hspace{4.5cm}(b)}\\
 \flushleft{\footnotesize {\bf Fig.~4.} $q_{0}=\sqrt{2}$, $\rho=3$, $\delta=\frac{\pi}{3}$, $\alpha=0$, $\eta=-\frac{\pi}{2}$. (a) the perspective view of one soliton. (b) the propagation view of soliton solution in different time.}
\end{center}
\subsection{Case II }
In this subsection, we assume the eigenvalues $\nu_{n}$ is the second kind eigenvalues. This implies that $N_{2}=1$ and $N_{1}=N_{3}=0$. And we also assume that
$\mathbf{q}_{+}=(1,1)^{T}\quad H_{1}=e^{\alpha+i\eta},\quad \zeta=i\rho e^{i\delta}$.
So we derived the soliton solution
\begin{equation}
q(x,t)=\frac{\det\left(
\begin{array}{ccc}
q_{+}&y_{1}&y_{2}\\
b_{1}&1-F_{11}&-F_{12}\\
b_{2}&-F_{21}&1-F_{22}
\end{array}
\right)}
{\det\left(
\begin{array}{cc}
1-F_{11}&-F_{12}\\
-F_{21}&1-F_{22}
\end{array}\right)},\nonumber
\end{equation}
where
\begin{subequations}
\begin{align}
&F_{11}=\Delta_{1}^{(9)}(\zeta_{1}),\quad F_{12}=\Delta_{1}^{(8)}(\zeta_{1}),\quad F_{21}=\Delta_{1}^{(4)}(\bar{\zeta}_{1}),\quad F_{22}=\Delta_{1}^{(5)}(\bar{\zeta}_{1}),\nonumber\\
&b_{1}=-\frac{iq_{+}}{\zeta_{1}},\quad b_{2}=-\frac{q_{+}}{q_{0}},\quad y_{1}=-\frac{\zeta_{1}}{q_{0}}\tilde{F}_{1}e^{2i\theta_{1}(-\frac{q_{0}^{2}}{\zeta_{1}})},\quad
y_{2}=-i\hat{F}_{1}e^{2i\theta_{1}(\bar{\zeta}_{1})},\nonumber\\
&\Delta_{1}^{(9)}(\zeta_{1})=\frac{\zeta_{1}}{iq_{0}}\frac{\tilde{F}_{1}e^{2i\theta_{1}(-\frac{q_{0}^{2}}{\zeta_{1}})}}{\zeta_{1}-(-\frac{q_{0}^{2}}{\zeta_{1}})},\quad
\Delta_{1}^{(8)}(\zeta_{1})=\frac{\hat{F}_{1}e^{2i\theta_{1}(\bar{\zeta}_{1})}}{\zeta_{1}-\bar{\zeta}_{1}},\nonumber\\
&\Delta_{1}^{(4)}(\bar{\zeta}_{1})=\frac{F_{1}e^{-2i\theta_{1}(\zeta_{1})}}{\bar{\zeta}-\zeta_{1}},\quad
\Delta_{1}^{(5)}(\bar{\zeta}_{1})=\frac{\bar{\zeta}_{1}}{iq_{0}}\frac{\check{F}_{1}e^{-2i\theta_{1}(-\frac{q_{0}^{2}}{\bar{\zeta}_{1}})}}{\bar{\zeta}_{1}-(-\frac{q_{0}^{2}}{\bar{\zeta}_{1}})}.\nonumber
\end{align}
\end{subequations}
Figure 5 showing the perspective view one soliton solution as $\rho>0$ and $\rho<0$. Figure 6 showing the propagation view of one soliton solution as different time.
\begin{center}
{\includegraphics[scale=0.45]{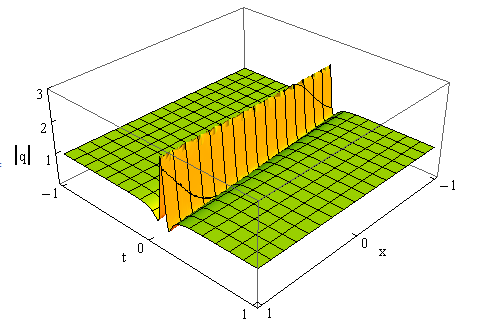}}\hspace{0.5cm}{\includegraphics[scale=0.35]{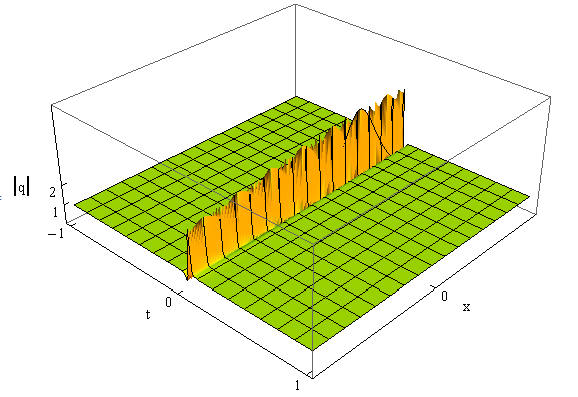}}\\
\vspace{-0.2cm}{\footnotesize\hspace{0.5cm}(a)\hspace{4.5cm}(b)}\\
 \flushleft{\footnotesize {\bf Fig.~5.} (a)the perspective view of one soliton solution as $\rho=2$, $\delta=\frac{\pi}{24}$.(b)the perspective view of one soliton solution $\rho=4$, $\delta=\frac{\pi}{24}$.}
\end{center}
\begin{center}
{\includegraphics[scale=0.35]{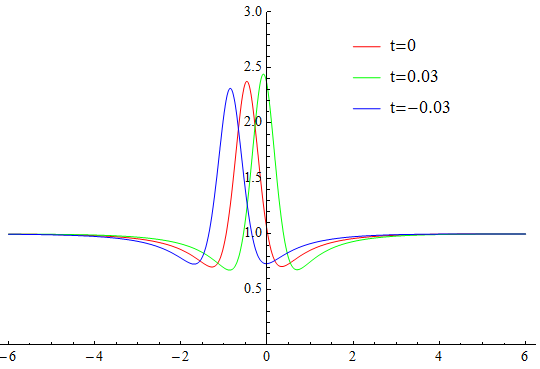}}\hspace{0.5cm}{\includegraphics[scale=0.35]{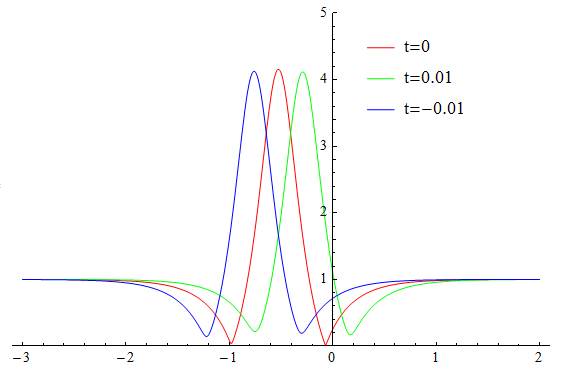}}\\
\vspace{-0.2cm}{\footnotesize\hspace{0.5cm}(a)\hspace{4.5cm}(b)}\\
 \flushleft{\footnotesize {\bf Fig.~6.} (a)the propagation view of one soliton solution as $\rho=2$, $\delta=\frac{\pi}{24}$.(b)the propagation view of soliton solution as $\rho=4$, $\delta=\frac{\pi}{24}$.}
\end{center}

\subsection{Case III  }
In this subsection, we assume the eigenvalues $\nu_{n}$ is the third kind eigenvalues. This implies that $N_{3}=1$ and $N_{1}=N_{2}=0$. And we also assume that
$\mathbf{q}_{+}=(1,1)^{T}\quad H_{1}=e^{\alpha+i\eta},\quad \omega_{1}=\rho e^{i\delta}$.
So we derived the soliton solution
\begin{equation}
q(x,t)=\frac{\det\left(
\begin{array}{cc}
q_{+}&y_{1}\\
b_{1}&1-F_{11}
\end{array}
\right)}{1-F_{11}},\nonumber
\end{equation}
where
\begin{subequations}
\begin{align}
&F_{11}=\Delta_{1}^{(2)}(\bar{\omega}_{1})\Delta_{1}^{(10)}(\omega_{1}),\quad b_{1}=-\frac{iq_{+}}{\omega_{1}}\Delta_{1}^{(2)}(\bar{\omega}_{1}),\quad
y_{1}=-i\hat{G}_{1}e^{i(\theta_{1}+\theta_{2})(\bar{\omega}_{1})},\nonumber\\
&\Delta_{1}^{(2)}(\bar{\omega}_{1})=\frac{G_{1}e^{-i(\theta_{1}+\theta_{2})(\omega_{n})}}{\bar{\omega}_{1}-\omega_{1}}+\frac{\omega_{1}}{iq_{0}}\frac{\tilde{G}_{1}e^{i(\theta_{1}-\theta_{2})(-\frac{q_{0}^{2}}{\omega_{1}})}}{\bar{\omega}_{1}-(-\frac{q_{0}^{2}}{\omega_{1}})},\nonumber\\
&\Delta_{1}^{10}(\omega_{1})=\frac{\hat{G}_{1}e^{i(\theta_{1}+\theta_{2})(\bar{\omega}_{1})}}{\omega_{1}-\bar{\omega}_{1}}.\nonumber
\end{align}
\end{subequations}
Figure 7 and Figure 8 showing the perspective view and propagation view of one soliton solution.
\begin{center}
{\includegraphics[scale=0.35]{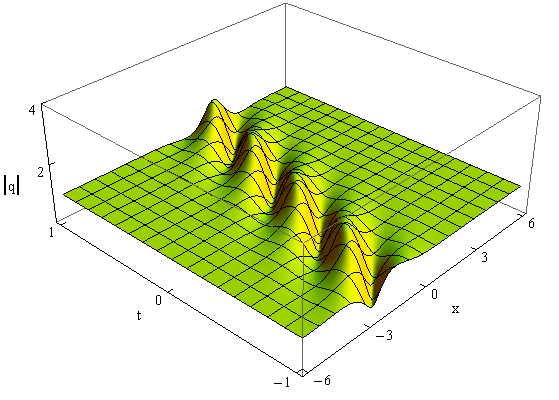}}\hspace{0.5cm}{\includegraphics[scale=0.35]{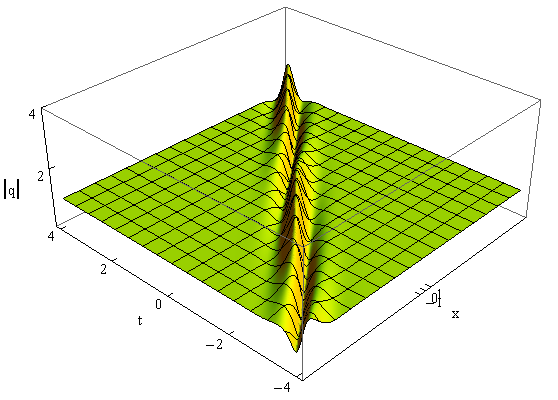}}\\
\vspace{-0.2cm}{\footnotesize\hspace{0.5cm}(a)\hspace{4.5cm}(b)}\\
 \flushleft{\footnotesize {\bf Fig.~7.} (a)the perspective view of one soliton solution as $\rho=2$, $\delta=\frac{\pi}{4}$.(b)the perspective view of soliton solution as $\rho=1$, $\delta=\frac{\pi}{4}$.}
\end{center}
\begin{center}
{\includegraphics[scale=0.35]{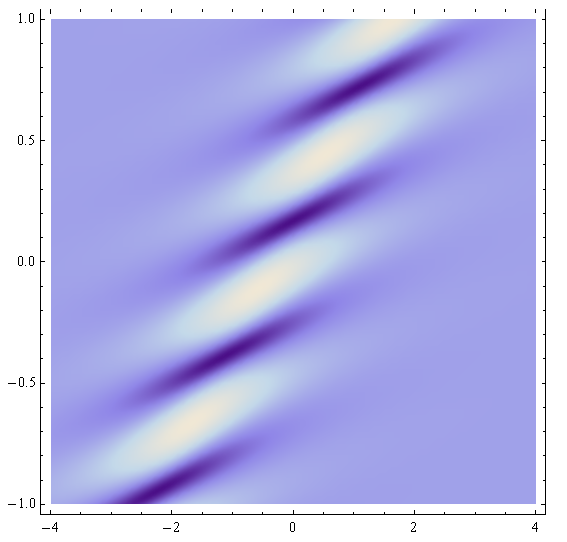}}\hspace{0.5cm}{\includegraphics[scale=0.35]{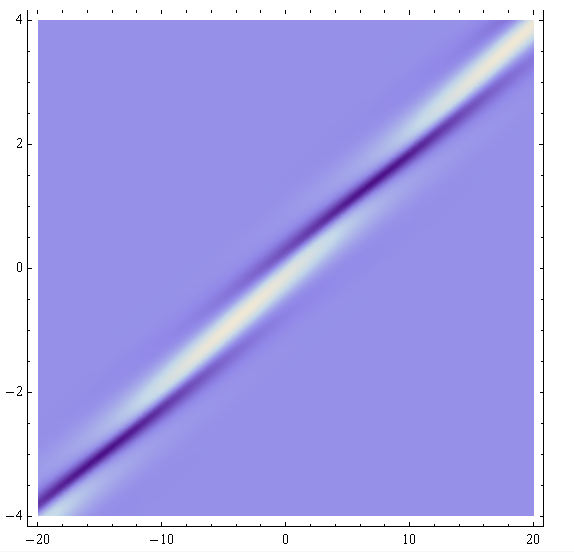}}\\
\vspace{-0.2cm}{\footnotesize\hspace{0.5cm}(a)\hspace{4.5cm}(b)}\\
 \flushleft{\footnotesize {\bf Fig.~8.} (a)the density view of one soliton solution as $\rho=2$, $\delta=\frac{\pi}{4}$.(b)the density view of soliton solution $\rho=1$, $\delta=\frac{\pi}{4}$.}
\end{center}

\section*{Acknowledgments}
 This work is supported by the National Science Foundation of China (Grant No.
11671095, 51879045).

\end{document}